\documentclass[preprint2,twocolappendix]{aastex6}

\usepackage{url}
\shorttitle{Disk--Jet Connection in Active Supermassive Black Holes}
\shortauthors{Inoue, Doi, Tanaka, Sikora, and Madejski}

\begin{document}
\title{Disk--Jet Connection in Active Supermassive Black Holes in the Standard Accretion Disk Regime} 

\author{Yoshiyuki Inoue\altaffilmark{1}, Akihiro Doi\altaffilmark{1}, Yasuyuki T. Tanaka\altaffilmark{2}, Marek Sikora\altaffilmark{3}, Gregorz M. Madejski\altaffilmark{4}} 

\affil{$^1$Institute of Space and Astronautical Science JAXA, 3-1-1 Yoshinodai, Chuo-ku, Sagamihara, Kanagawa 252-5210, Japan}
\affil{$^2$Hiroshima Astrophysical Science Center, Hiroshima University, 1-3-1 Kagamiyama, Higashi-Hiroshima, Hiroshima 739-8526, Japan}
\affil{$^3$Nicolaus Copernicus Astronomical Center, Bartycka 18, 00-716 Warsaw, Poland}
\affil{$^4$Kavli Institute for Particle Astrophysics and Cosmology, SLAC National Accelerator Laboratory, Stanford University, 2575 Sand Hill Road M/S 29, Menlo Park, CA 94025, USA}

\altaffiltext{1}{e-mail: yinoue@astro.isas.jaxa.jp}

\begin{abstract}
We study the disk-jet connection in supermassive black holes by investigating the properties of their optical and radio emissions utilizing the SDSS-DR7 and the NVSS catalogs. Our sample contains 7017 radio-loud quasars with detection {\it both} at 1.4~GHz and SDSS optical spectrum. Using this radio-loud quasar sample, we investigate the correlation among the jet power ($P_{\rm jet}$), the bolometric disk luminosity ($L_{\rm disk}$), and the black hole mass ($M_{\rm BH}$) in the standard accretion disk regime. We find that the jet powers correlate with the bolometric disk luminosities as $\log P_{\rm jet} = (0.96\pm0.012)\log L_{\rm disk} + (0.79 \pm 0.55)$. This suggests that the jet production efficiency of $\eta_{\rm jet}\simeq1.1_{-0.76}^{+2.6}\times10^{-2}$ assuming the disk radiative efficiency of $0.1$ implying low black hole spin parameters and/or low magnetic flux for radio-loud quasars. But it can be also due to dependence of the efficiency on geometrical thickness of the accretion flow which is expected to be small for quasars accreting at the disk Eddington ratios $0.01 \lesssim \lambda \lesssim 0.3$. This low jet production efficiency does not significantly increase even if we set the disk radiative efficiency of $0.3$. We also investigate the fundamental plane in our samples among $P_{\rm jet}$, $L_{\rm disk}$, and $M_{\rm BH}$. We could not find a statistically significant fundamental plane for radio-loud quasars in the standard accretion regime.
\end{abstract}

\keywords{accretion, accretion disks - black hole physics  - galaxies: active - galaxies: jets - quasars: supermassive black holes}

\section{Introduction}
\label{sec:intro}
Relativistic jets launched by supermassive black holes (SMBHs), so-called as active galactic nuclei (AGNs), are known as the most energetic particle accelerators in the universe. Because of their gigantic power, those jets would affect the fate of galaxies, and also galaxy clusters \citep[e.g.][]{fab12}. However, the launching mechanism of a collimated relativistic jet from a black hole system is a long standing problem in astrophysics. 

Theoretically, the Blandford-Znajek (BZ) mechanism \citep{bla77} is believed as the plausible explanation for the jet launch. In the BZ mechanism, the jet power is extracted by the rotation of BHs with the support of the magnetic fields threading the central BH. Recent numerical simulations confirm this process as a plausible and efficient jet power extraction mechanism \citep[e.g.][]{kom07,tch10,tch11,mck12,tak16}. 

Observational evidence for the jet production mechanisms in AGNs are not clear yet. A possible and important key to understand the jet launching mechanism is to measure spins of the nuclear BHs which is a key parameter for the jet. For example, \citet{sik07} argued that spin parameter would determine the AGN radio loudness distribution. Although various spin measurement methods have been proposed in literature \citep[e.g.][and references therein]{mor15}, spin determination of active SMBHs is currently uncertain \citep[e.g.][]{mar11,bre11,kin13,liu15}. 

It is also possible to investigate the jet production mechanisms from observations by probing the relation between disk inflow and jet outflow, because a part of infall materials to the central BH are ejected as the jet outflow and those accreting material also accumulate the magnetic field in the vicinity of the SMBHs. For example, recent systematic spectral analysis for luminous blazars by \citet{ghi14} shows the jet power is slightly larger than the power of accreting plasma, although the results depend on the assumptions on the pair fraction and the minimum energy of elections \citep[e.g.][]{ino16,pja16}. On the contrary to blazars, in radio quasars, the efficiency of jet energy extraction from rest energy of accreted masses, i.e. the jet production efficiency, is not well understood yet.

Here, after the discovery of quasars, their radio emission is found to be associated with a presence of luminous optical emission lines \citep{baa54,ost77,gra78}. A correlation between radio jet luminosities and disk emission line luminosities in radio galaxies is later found \citep[e.g.][]{bau89,sau89,raw89,raw91,zir95,wil99,but10,koz11,sik13}. The correlation of radio jet luminosities and various optical emissions has been studied in more detail as the mass of the nuclear BH became available \citep[e.g.][]{woo02}. The BH mass provide fundamental information for the jet study such as the Eddington luminosity and its ratio.

Using the BH mass information, BH systems have been found to have a fundamental plane among mass, disk luminosity, and jet power \citep[e.g.][]{ter03, mer03,mac03}. \citet{mer03} established the fundamental plane in stellar mass and supermassive BHs using BH mass, jet core radio luminosity, and disk X-ray luminosity. With $\sim150$ BH system samples, they found that the radio core luminosity is correlated with both the mass and the X-ray luminosity of the disk. However, AGN disk luminosity is known to be dominated not in X-ray but in optical \citep[e.g.,][]{elv94}. X-ray emission is reprocessed optical emission via Comptonization processes in hot accretion disk coronae \citep[e.g.,][]{kat76,poz77,sun80}. Moreover, as recent studies suggest that the fundamental plane exist only for low accretion rate BH systems \citep[see e.g.][]{mer08,plo12}, a fundamental plane for objects accreting at high rates is not well established.

The Sloan Digital Sky Survey \citep[SDSS;][]{yor00} has facilitated studies of AGNs in optical with wide and deep field surveys \citep[e.g.][]{she11}. SDSS has detected 166583 quasars in optical covering about a quarter of the sky \citep{par14}. For the radio data, using Very Large Array (VLA), the NRAO VLA Sky Survey (NVSS) has observed the entire sky north of $-40$~deg declination at 1.4~GHz down to $\sim2.5$~mJy \citep{con98}. NVSS has detected $\gtrsim$1.8~million sources. In this paper, we study the relation among the jet power ($P_{\rm jet}$), the bolometric disk luminosity ($L_{\rm disk}$), and the black hole mass ($M_{\rm BH}$) using quasar samples detected {\it both} in SDSS and NVSS which will allow us to investigate the disk--jet connection in the largest ever radio quasar sample.

This paper is organized as follows. In Section \ref{sec:sample}, we introduce the sample used in our analysis. In Section \ref{sec:jet-disk}, the relation between accretion inflows and jet outflows is presented. SMBH fundamental planes are discussed in Section \ref{sec:plane}. Discussion and conclusion is given in Section \ref{sec:dis} and Section \ref{sec:con}, respectively. Throughout this paper, we adopt the standard cosmological parameters of $(h, \Omega_M , \Omega_\Lambda) = (0.7, 0.3, 0.7)$.

\section{Sample}
\label{sec:sample}

The NVSS was carried out utilizing the VLA radio interferometric telescopes at a frequency of 1.4~GHz \citep{con98}. The NVSS was conducted with the array D configuration which provides a spatial resolution of 45~arcsec (corresponding to the physical size of $\sim76$~kpc at $z=1$). Although this resolution is not as good as that of the Faint Images of the Radio Sky at Twenty-cm \citep[FIRST;][]{bec95} having $\sim5$~arcsec (corresponding to the size of $\sim8.5$~kpc at $z=1$), the NVSS resolution secures to measure fluxes of extended sources more accurately \citep{lu07}. 

The NVSS covers the entire sky north of $-40$~deg declination ($\sim33,000~{\rm deg}^2$) and contains over 1.8~million sources down to a limiting flux density of $\sim2.5$~mJy. The survey gives astrometric accuracy ranges from 1~arcsec for bright sources to $\sim$7~arcsec for faint sources. We extract integrated flux densities from the NVSS catalog to evaluate the whole extended radio flux.

The SDSS is an optical imaging and spectroscopic survey \citep[e.g.,][]{yor00} using a 2.5~m wide-filed telescope at the Apache Point Observatory \citep{gun06}. The photometric survey contains five wavelength bands \citep[{\it ugriz};][]{fuk96}. The survey covers about a quarter of the sky. A subset of photometric sources are chosen for spectroscopic observation according to the spectral target selection algorithms of SDSS. The survey catalog now contains 166583  quasars \citep{par14}. Here, \citet{she11} provided detailed spectral properties of the SDSS data release (DR) 7 quasar catalog \citep{sch10} which contains 105783 quasars brighter than $M_i=-22.0$. In this paper, we utilize the SDSS DR7 quasar catalog provided by \citet{she11} as we are interested in the spectral properties of quasars. 

From the SDSS-DR7 quasar catalog, we extract redshifts, rest-frame 2500~\AA~fluxes, bolometric luminosities, disk Eddington ratios, and virial BH mass estimates. 2500~\AA~fluxes at the rest frame were determined from the power-law continuum fit to the spectrum. Bolometric luminosities were computed from continuum luminosities at 5100~\AA~($z<0.7$), 3000~\AA~($0.7\le z<1.9$), and 1350~\AA~($z\ge1.9$) using the bolometric luminosity correction factors of 9.26, 5.15, and 3.81 from the spectral energy distribution (SED) templates in \citet{ric06}. Although various studies report different correction factors \citep[see e.g.][]{elv94,ric06,nem10,run12,kra13}, those studies are consistent with in a factor of $\sim2$. We note that this bolometric luminosity includes emission from dust (infrared), accretion disk (optical), and corona (X-ray). In this paper, we approximate this bolometric luminosity equals to the bolometric disk luminosity, since SDSS quasars have $L_{\rm disk, 2500\AA}\gtrsim10^{44}~{\rm erg~s^{-1}}$ where bolometric luminosities are dominated by accretion disks. 

 BH masses were estimated from single-epoch spectra (virial mass) in the SDSS quasar survey \citep{she11}\footnote{Spatially resolved kinematics observations are limited to only nearby sources. Various indirect mass measurement methods have been developed such as the reverberation mapping \citep[e.g.][]{bla82,pet93,kas00} and the correlation between the optical luminosity and the broad-line-region size \citep[e.g.][]{kas00,mcl01,ves02}. }. Using the continuum luminosity and full width at half-maximum (FWHM), the virial mass estimate is given by
\begin{equation}
\log \left(\frac{M_{\rm BH}}{M_\odot}\right) = a + b \log \left(\frac{\lambda L_\lambda}{10^{44}~{\rm erg\ s}^{-1}}\right)+2\log \left(\frac{{\rm FWHM}}{{\rm km~s}^{-1}}\right),
\end{equation}
where $a$ and $b$ are the empirical coefficients calibrated with local AGNs using the reverberation mapping method \citep[e.g.][]{bla82,pet93,kas00}. \citet{she11} use the line of H$\beta$ for $z<0.7$, Mg$_{\rm II}$ for $0.7\le z<1.9$, and C$_{\rm IV}$ for $z\ge1.9$. The coefficients are as follows: $(a, b) = (0.910,0.50), (0.740,0.62)$, and $(0.660,0.53)$ for H$\beta$ \citep{ves06}, Mg$_{\rm II}$ \citep{mcl04,she11}, and C$_{\rm IV}$ \citep{ves06}, respectively.

\citet{she11} matched the DR7 quasar catalog with the FIRST catalog with a matching radius of 30~arcsec. In this paper, we match the NVSS radio sources and the DR7 quasar catalog with a matching radius of 30~arcsec and select the closest sources between the catalogs. As we are targeting SDSS detected quasars, the dominant radio quasar population is expected to be Fanaroff-Riley Class II \citep[FR-II;][]{fan74} whose radio brightest component, so-called hot spot, locates within several hundreds kpc away from the nucleus \citep[e.g.,][]{mul08} which corresponds to $\sim30$~arcsec at $z\sim$1--3. 

\citet{bes05} matched the SDSS DR2 galaxy catalog and the NVSS catalog out to 3~arcmin. They found that positional offsets between SDSS galaxies and their nearest NVSS source is significant out to 100~arcsec, although they concluded that true associations is expected to be smaller than 15~arcsec. We consider the case for a matching radius of 15~arcsec later at \autoref{subsec:offset}. Our results do not change even if adopting different matching radiuses.

\begin{figure}[t]
 \begin{center}
  \includegraphics[bb=0 0 864 648,width=8.5cm]{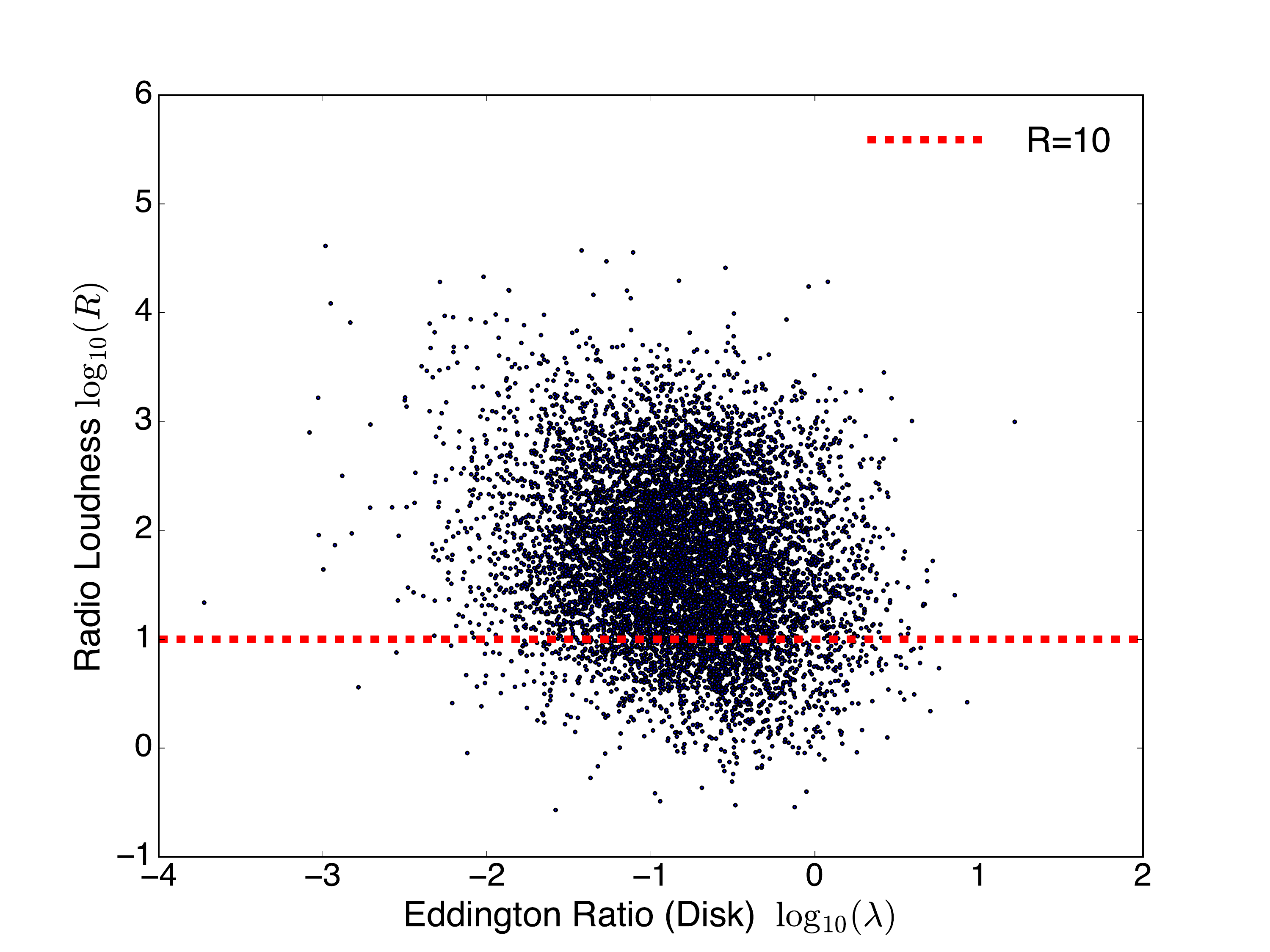} 
 \end{center}
\caption{Accretion disk Eddington ratio $\lambda$ vs. radio loudness $R$ of our SDSS--NVSS quasar sample. $R=10$ is shown by the dashed line.}\label{fig:lambda-R}
\end{figure}

After the catalog matching, the resulting number of objects detected both in radio and optical is 8436. To examine a luminosity correlation in radio galaxies, we derive the rest-frame radio and optical luminosities as follows. The rest luminosity at a frequency $\nu_0$ in the unit of ${\rm erg\ s^{-1}\ Hz^{-1}}$ is obtained as
\begin{equation}
L_{\nu} (\nu_0) = 4\pi d_L(z)^2 (1+z)^{\alpha-1} F_{\nu}(\nu_0),
\end{equation}
where $d_L(z)$ is the luminosity distance at a redshift $z$, $\alpha$ is the spectral index, and $F_\nu$ is the observed flux. For the radio spectral index (i.e., $F_\nu\propto\nu^{-\alpha}$ in the unit of mJy), we assume $\alpha=0.8$ \citep{kim08,sik13}. The SDSS DR7 catalog provides the rest-frame flux $F_{\lambda, {\rm rest}}$ in the unit of $[{\rm erg~cm^{-2}~s^{-1}~\AA^{-1}}]$. \autoref{fig:lambda-R} shows the distribution of SDSS-NVSS quasars in the space of the accretion disk Eddington ratio $\lambda\equiv L_{\rm disk}/L_{\rm Edd}$ and the radio loudness $R\equiv L_{\rm 5~GHz}/L_{B-\rm band}$ \citep[e.g.][]{sik07}. For the optical spectral index, we assume $\alpha=0.5$ \citep{ric06}.

\begin{figure}
 \begin{center}
  \includegraphics[bb=0 0 864 648,width=8.5cm]{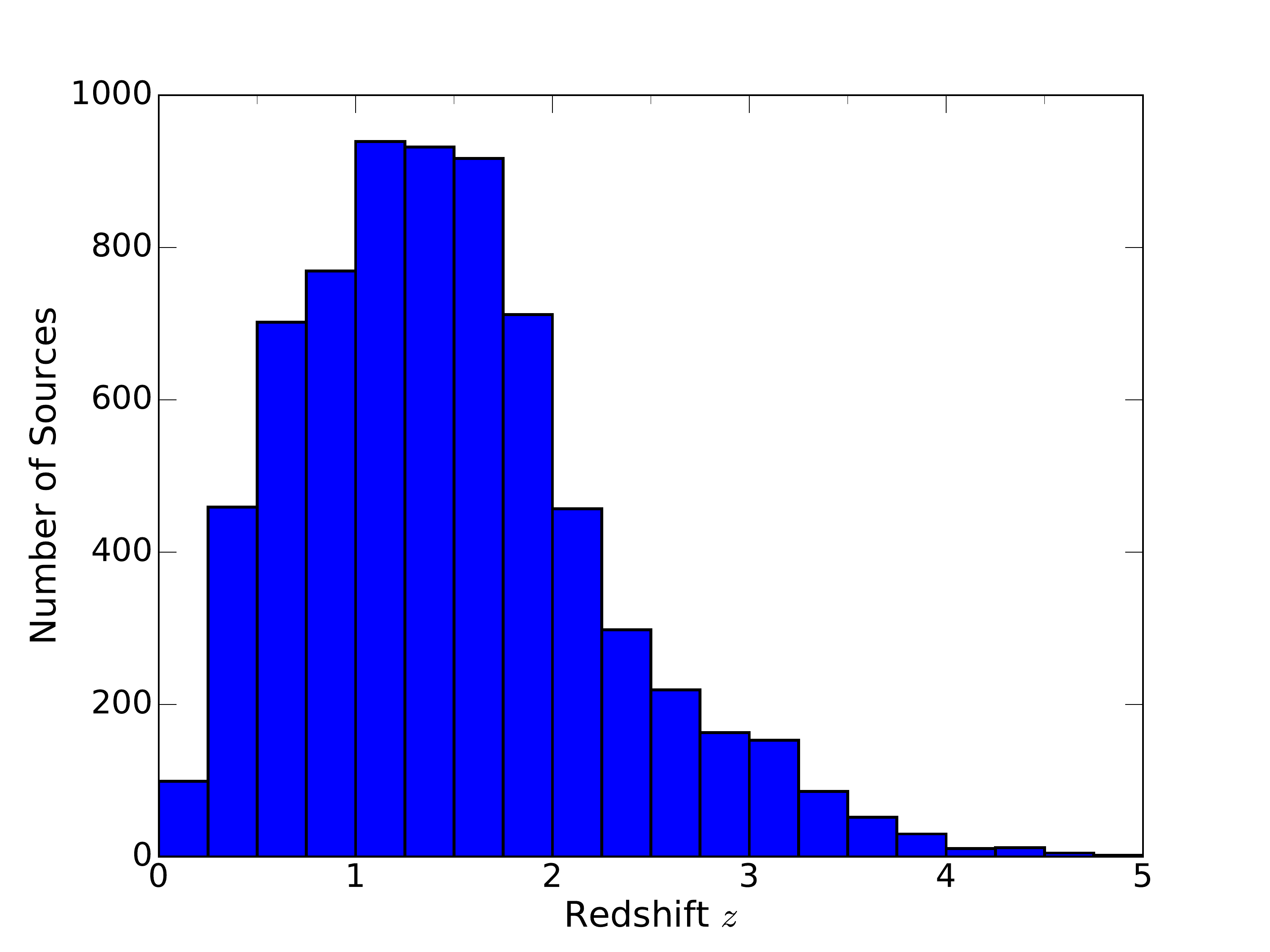} 
 \end{center}
\caption{Redshift distribution of the radio-loud SDSS-NVSS quasar sample. 7017 objects are contained.}\label{fig:redshift}
\end{figure}

We are interested in the relation between the jet and the accretion disk, i.e. radio-loud objects. Radio emission from radio-quiet quasars ($R<10$) is expected to be associated with shocks produced by  quasar driven outflows rather than jets \citep[e.g.][]{zak14}. To select radio-loud quasars, we select sources having $R\geq10$ for the main sample which includes 7017 quasars. The redshift distribution of our radio-loud quasar sample is shown in \autoref{fig:redshift}. Redshifts of our sample range from 0.077 to 4.922. Majority of the samples ($\sim$87\%) are at $z\le2.5$.

\begin{figure}[t]
 \begin{center}
  \includegraphics[bb=0 0 864 648,width=8.5cm]{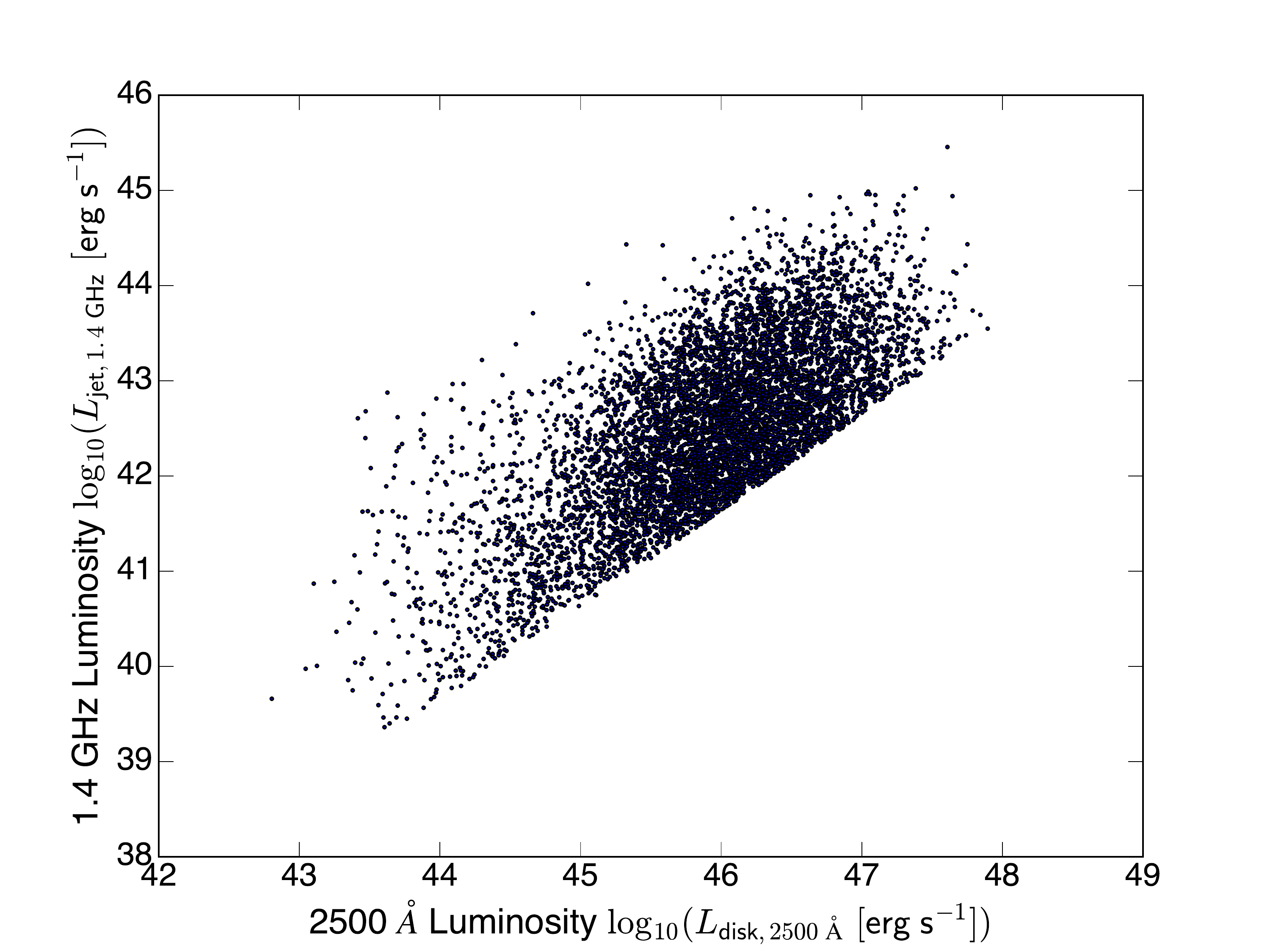} 
 \end{center}
\caption{Optical luminosity at 2500~\AA~vs. radio luminosity at 1.4~GHz of our radio-loud SDSS--NVSS quasar sample ($R\geq10$). }\label{fig:Ldisk_Ljet}
\end{figure}

\autoref{fig:Ldisk_Ljet} shows the 2500~\AA~and 1.4~GHz luminosity relation of our SDSS--NVSS radio-loud quasar samples. We need to examine the reliability of the correlation between the optical and radio luminosities. In the flux-limited samples, luminosities of samples can be strongly correlated with redshifts due to the detection limits. This might result in a spurious correlation. To avoid this, we perform a partial correlation analysis \citep[see e.g.][]{pad92,ghi11,ino11} to test the correlation between the logarithmic optical and radio luminosities excluding the redshift dependence. The Spearman rank correlation gives $\rho_{{\rm or},z}=0.39$ with the p-value of $<10^{-10}$ (i.e. the probability of the null hypothesis). Therefore, there is a weak positive correlation between optical and radio luminosities in our radio quasar samples. The correlation analysis results in this paper are summarized in \autoref{tab:corr}. The tests are performed using the astronomy survival analysis code ASURV \citep{lav92} implemented in the STSDAS package in iraf. 

\floattable
\begin{deluxetable}{lcccccc}
 \tablecaption{Results of Correlation Analysis\label{tab:corr}}
\tablehead{
  \colhead{x} & \colhead{y}  & \colhead{z} & \colhead{Objects} & \colhead{Samples} & \colhead{$\rho_{\rm xy,z}$\tablenotemark{a}} & \colhead{${\mathrm p}$-value}\
 }
\startdata
$\log L_{\rm 2500~\AA}$ & $\log L_{\rm 1.4~GHz}$ & $z$ & SDSS-NVSS quasars\tablenotemark{b} & 7017 & 0.39  &  $<10^{-10}$ \\
$\log L_{\rm disk}$ & $\log P_{\rm jet}$ & $z$ & SDSS-NVSS quasars\tablenotemark{b}  & 7017 & 0.40  & $<10^{-10}$\\
$\log L_{\rm disk}$ & $\log P_{\rm jet}$ & $\log M_{\rm BH}$ & SDSS-NVSS quasars\tablenotemark{b} & 7017 & 0.63  & $<10^{-10}$\\
$\log M_{\rm BH}$ & $\log L_{\rm disk}$ & $\log P_{\rm jet}$ & SDSS-NVSS quasars\tablenotemark{b} & 7017 & 0.37 & $<10^{-10}$\\
$\log M_{\rm BH}$ & $\log P_{\rm jet}$ & $\log L_{\rm disk}$ & SDSS-NVSS quasars\tablenotemark{b} & 7017 & -0.028  & 0.019\\
$\log \lambda$  & $\log r$ & - & SDSS-NVSS quasars\tablenotemark{c} & 8436 & -0.11  & $<10^{-10}$\\
$\log \lambda$  & $\log r$ & $\log L_{\rm disk}$ & SDSS-NVSS quasars\tablenotemark{c} & 8436 & -0.022  & 0.065\\
$\log L_{\rm disk}$ & $\log P_{\rm jet}$ & $z$ & SDSS-NVSS-WENSS quasars\tablenotemark{b}  & 2077 & 0.31  & $<10^{-10}$\\
$\log L_{\rm disk}$ & $\log P_{\rm jet}$ & $z$ & uniform SDSS-NVSS quasars\tablenotemark{b}  & 3545 & 0.37  & $<10^{-10}$\\
$\log L_{\rm disk}$ & $\log P_{\rm jet}$ & $z$ & SDSS-NVSS quasars\tablenotemark{d}  & 6081 & 0.39  & $<10^{-10}$\\
$\log L_{\rm disk}$ & $\log P_{\rm jet}$ & $z$ & SDSS-NVSS quasars\tablenotemark{e}  & 3083 & 0.63  & $<10^{-10}$\\
\enddata
\tablenotetext{a}{Spearman's rank correlation coefficient.}
\tablenotetext{b}{Radio-loud objects only, $R\geq10$.}
\tablenotetext{c}{All the SDSS-NVSS quasars are included.}
\tablenotetext{d}{The matching radius is set to be 15~arcsec.}
\tablenotetext{e}{$R\geq100$.}
\end{deluxetable}

\begin{figure*}[t]
 \begin{center}
  \includegraphics[bb=0 0 864 648,width=15.0cm]{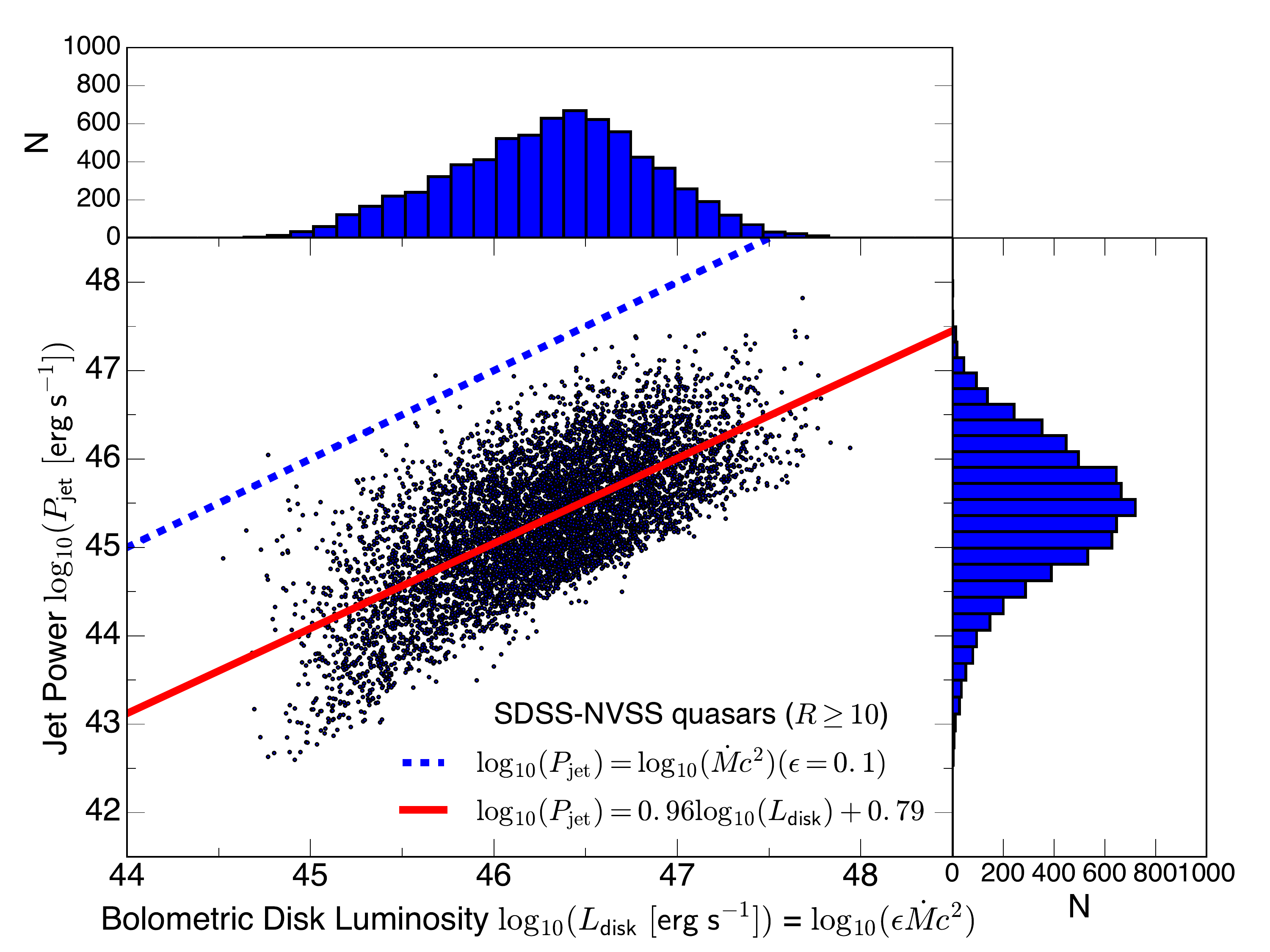} 
 \end{center}
\caption{Bolometric disk luminosity vs. jet power  of our SDSS--NVSS sample, where the top and right panels show the bolometric disk luminosity and jet power histograms. The solid line gives a linear fit to the data (see \autoref{eq:jet_disk}) with a scatter of 0.74. The dashed line shows the case with the jet production efficiency $\eta_{\rm jet}=1$ assuming the disk efficiency $\epsilon=0.1$.}\label{fig:Ldisk_Pjet}
\end{figure*}

\section{Relation between Jet Outflow and Accretion Disk Inflow}
\label{sec:jet-disk}

The jet power is known to be correlated with the radio luminosity \citep[e.g.][]{wil99,osu11}. The empirical relation gives the time-averaged jet power from the extended radio luminosity as \citep{wil99}
\begin{equation}
P_{\rm jet} = 9.5\times10^{46} \left(\frac{f}{10}\right)^{3/2} \left(\frac{L_{\rm 151~MHz}}{10^{28}~{\rm W~Hz^{-1}~sr^{-1}}}\right)^{6/7}~[{\rm erg~s^{-1}}],
\label{eq:Pjet_wil99}
\end{equation}
where $f$ is a parameter accounting for systematic error in the model assumptions. The following parameters are absorbed in the parameter $f$: the factor accounting for energy loss via the adiabatic expansion, the factor accounting for the bulk and turbulent kinetic energy of the lobe, the energy fraction of radiating particles, the angle between the magnetic field direction and the line-of-sight, the low frequency cutoff in the synchrotron spectrum, the volume filling factor, and deviations from the minimum-energy condition. \citet{wil99} constrained as $1\le f \lesssim 20$ by using X-ray cavity measurements. In this paper, we set $f=10$ following the recent X-ray cavity and hot spot studies \citep[][]{blu00,god13}. We note the power given in \autoref{eq:Pjet_wil99} is the time-averaged value.

As we use the NVSS detected samples with angular resolution of 45~arcsec, there are unresolved sources in our catalog. \citet{sha13} have recently investigated the effect of source size in the jet power estimation of \autoref{eq:Pjet_wil99}. They found the dependence on the source size is the power of $0.58\pm0.17$. Thus, the effect of the source size is expected to be small in the jet power estimation, although it might affect the power estimates for ultra-compact or largely extended sources. 

Since our samples are quasars, $L_{\rm disk}$ can be related to the mass accretion rate $\dot{M}_{\rm in}$ onto the SMBH as
\begin{equation}
L_{\rm disk} = \epsilon \dot{M}_{\rm in} c^2,
\end{equation}
where $\epsilon$ is the accretion disk radiative efficiency and $c$ is the speed of light. \autoref{fig:Ldisk_Pjet} shows the correlation between the bolometric disk luminosity $L_{\rm disk}$ and the jet power $P_{\rm jet}$ in the logarithmic space. The top and bottom panel shows the histogram of $\log (L_{\rm disk})$ and $\log(P_{\rm jet})$, respectively. The linear regression line, shown by the solid line, is given as
\begin{equation}
\log P_{\rm jet} = (0.96\pm0.012)\log L_{\rm disk} + (0.79 \pm 0.55),
\label{eq:jet_disk}
\end{equation}
where errors show $1\sigma$ uncertainties, with a scatter of 0.54. We perform a partial correlation analysis to test the correlation between the $\log L_{\rm disk}$ and $\log P_{\rm jet}$ excluding the redshift dependence (\autoref{tab:corr}). The Spearman rank correlation gives $\rho_{LP,z}=0.40$ with the p-value of $<10^{-10}$. Therefore, there is a positive correlation between disk luminosity and jet power in radio quasars. We note that, in this paper, we do not include radio non-detected data in the analysis. The effect of such censored data is discussed in \autoref{app:corr}.

\begin{figure}[t]
 \begin{center}
  \includegraphics[bb=0 0 864 648,width=8.5cm]{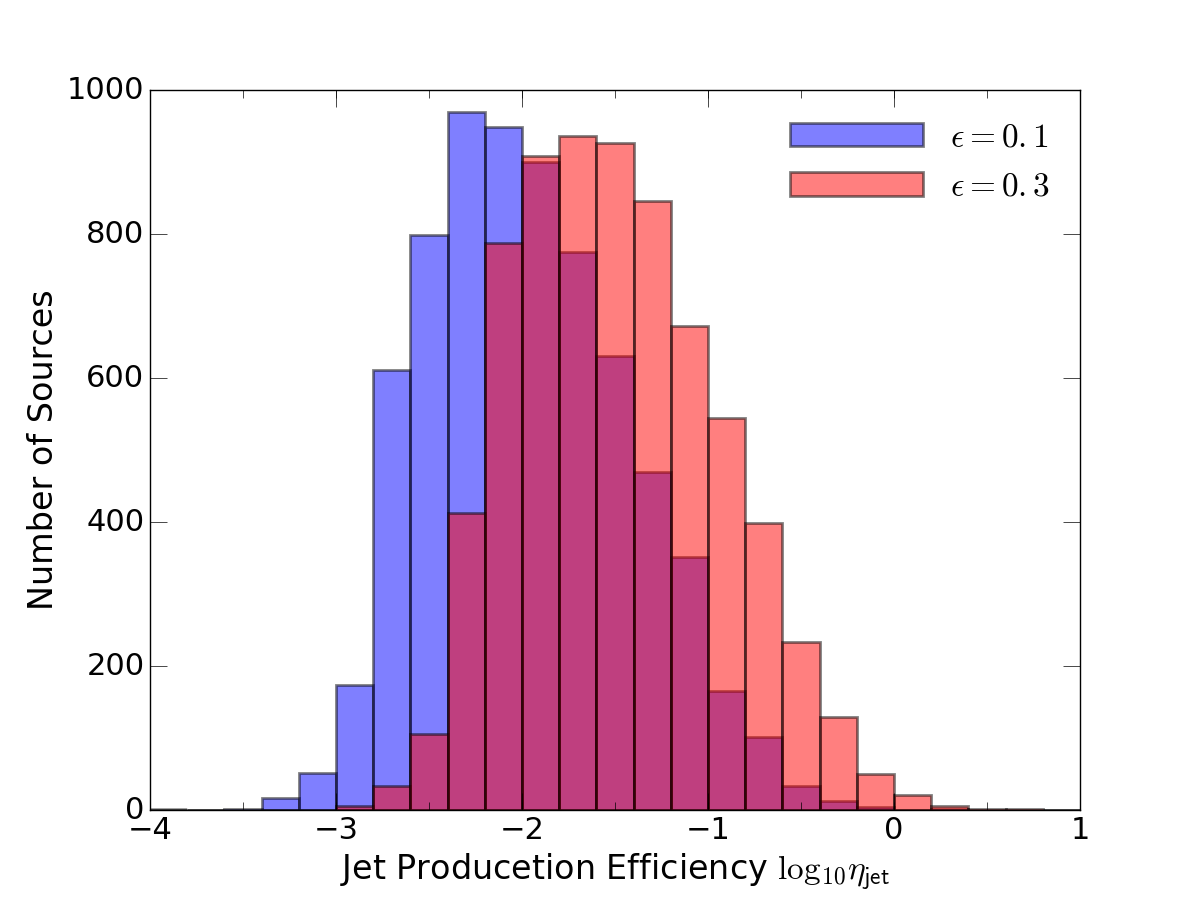} 
 \end{center}
\caption{Distribution of the jet production efficiencies of our radio quasar samples. We assume the radiative efficiency of the disk $\epsilon$ to be 0.1 (blue) and 0.3 (red).}\label{fig:hist_logeta}
\end{figure}

The jet production efficiency is defined as
\begin{equation}
\eta_{\rm jet} \equiv \frac{P_{\rm jet}}{\dot{M}_{\rm in} c^2} = \frac{P_{\rm jet}}{L_{\rm disk}/\epsilon}.
\end{equation} 
The dashed line in \autoref{fig:Ldisk_Pjet} represents $P_{\rm jet} = L_{\rm disk}/\epsilon=\dot{M}_{\rm in} c^2$ with $\epsilon = 0.1$ corresponding to $\eta_{\rm jet}=1$. The distribution of the jet production efficiency of our radio quasar samples are shown in \autoref{fig:hist_logeta}. We show the cases of $\epsilon=0.1$ and $0.3$, although the efficiency can be from 0.057 for a Schwarzschild BH to 0.42 for an extreme Kerr BH \citep[e.g.][]{kat98}. Here, $<\log \eta_{\rm jet}>=-1.97\pm0.54$ corresponding to $\eta_{\rm jet}\simeq1.1_{-0.76}^{+2.6}\times10^{-2}$ for $\epsilon=0.1$. Only 2 quasars have $\eta_{\rm jet}\geq1$. This low jet production efficiency is consistent with the study by \citet{van13} in which they used a correlation between optical and radio luminosities of 763 FR-II quasars utilizing the SDSS-FIRST catalog.  This low efficiency is not significantly enhanced even if we set $\epsilon=0.3$ (\autoref{fig:hist_logeta}). 

Connection between the spin value and radiative efficiency is theoretically established for standard accretion disk. It comes from energetics of particles on marginally stable orbits and gives efficiency of radiation production. At large spins, a fraction of radiation is captured by the central BH itself. In this case, the observed efficiency would become lower. Furthermore, innermost portions of the disk is dominated by magnetic fields in the magnetically arrested disk \citep[MAD;][]{nar03} framework which is the current leading scenario for the powerful jet launching mechanism (see \autoref{subsec:impjet} for details). Although their radiative efficiency is expected to be different from that for standard disks, the exact value of the efficiency is still under debate \citep[see e.g.,][]{pun14,pun15,pun16,ava16}. Considering these uncertainties, we fix $\epsilon=0.1$ in this paper.

\section{Fundamental Planes in Supermassive Black Holes}
\label{sec:plane}

Low-accretion rate BH systems ($\lambda\lesssim0.03$) have been believed to have a fundamental plane among mass, disk luminosity, and jet power \citep[e.g.][]{ter03, mer03,mac03,plo12}, while the existence of a fundamental plane for objects accreting at high rates is not well established. 

\citet{mer03} established a fundamental plane in stellar mass and supermassive black hole systems using mass, radio luminosity, and X-ray luminosity. With $\sim150$ samples, they found that the radio luminosity is correlated with both the BH mass and the X-ray luminosity. However, the SMBH disk luminosity is dominated in optical bands rather than in X-rays \citep{elv94}. And, they used the radio luminosity of the core component only rather than the extended component which includes the whole energy budget of the jet. In this section, we examine the fundamental plane in the SMBH systems with our 7017 radio-loud quasar samples having $\lambda\gtrsim0.01$.

We examine the mass dependence of the correlation between $L_{\rm disk}$ and $P_{\rm jet}$ (\autoref{fig:Ldisk_Pjet}). The Spearman rank correlation gives $\rho_{LP, M}=0.63$ with the p-value of $<10^{-10}$. Therefore, the positive correlation between disk luminosity and jet power still exist, even if we consider the dependence on the BH mass. 

The correlations between $M_{\rm BH}$ and $L_{\rm disk}$ or $P_{\rm jet}$ are also examined. For the correlation between $M_{\rm BH}$ and $L_{\rm disk}$, the Spearman rank correlation also gives $\rho_{ML, P}=0.37$ with the p-value of $<10^{-10}$. On the contrary, for the correlation between $M_{\rm BH}$ and $P_{\rm jet}$, the Spearman rank correlation gives $\rho_{MP, L}=-0.028$ with the p-value of $0.019$. Thus, once we exclude the dependence on $L_{\rm disk}$, there is no correlation between $M_{\rm BH}$ and $P_{\rm jet}$ in radio-loud luminous quasars. Therefore, the disk luminosity correlates with both the BH mass and the jet power, while the jet power correlates with the disk luminosity, i.e. absolute mass accretion rate, but not with the BH mass. This result is consistent with semi-analytical studies of SMBH growth \citep{mer08}.

The multivariate regression analysis can give the following forms of fitting equations
\begin{eqnarray}
\log L_{\rm disk} &=& (23.8\pm0.23) + (0.43\pm5.5\times10^{-3}) \log P_{\rm jet} \nonumber \\
&&+ (0.33\pm9.4\times10^{-3}) \log M_{\rm BH}, \label{eq:plane_LPM}\\
\log P_{\rm jet} &=& (7.4\pm0.54) + (0.84\pm1.3\times10^{-2}) \log L_{\rm disk} \nonumber \\
&&+ (-0.057\pm1.5\times10^{-2}) \log M_{\rm BH}. \label{eq:plane_PLM}
\end{eqnarray}
The scatter of each function is 0.37 and 0.59 for \autoref{eq:plane_LPM} and \autoref{eq:plane_PLM}, respectively. Here, we note that the correlation between $P_{\rm jet}$ and $M_{\rm BH}$ is not statistically established as discussed above. And, the slope coefficient of $\log M_{\rm BH}$ term is small as $-0.057$ implying very weak dependence. 

\section{Discussions}
\label{sec:dis}

\subsection{Effect of Spectral Index}

\begin{figure}[t]
 \begin{center}
  \includegraphics[bb=0 0 864 648,width=8.5cm]{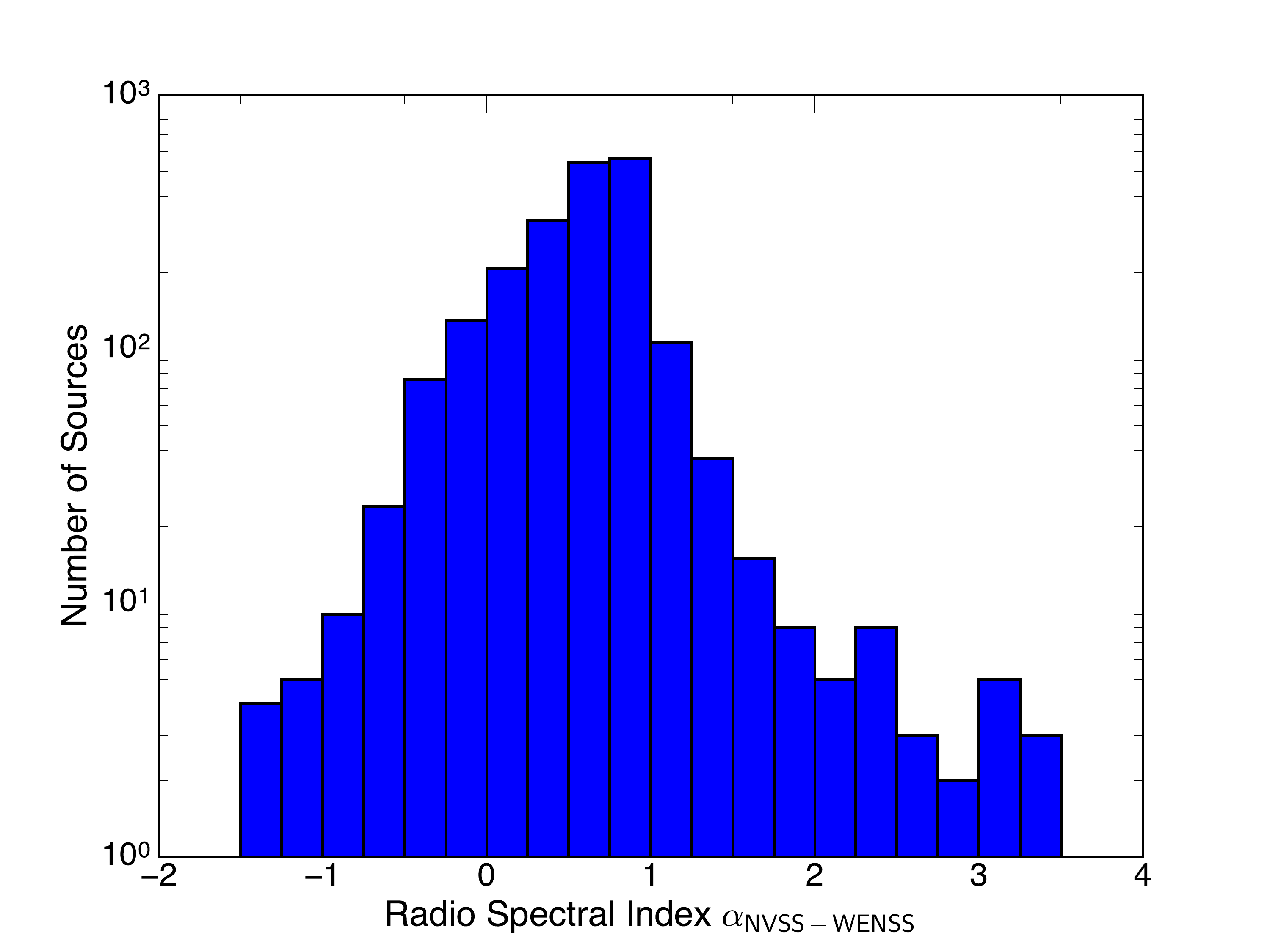} 
 \end{center}
\caption{Distribution of the radio spectral index $\alpha_{\rm NVSS-WENSS}$ between 1.4~GHz (NVSS) and 325~MHz (WENSS) of our SDSS-NVSS-WENSS detected quasar sample.}\label{fig:alpha_WENSS}
\end{figure}

In this paper, we have assumed that the radio spectral index is uniformly $\alpha=0.8$. However, individual sources would have different spectral indices, which might affect the results \citep[e.g.,][for the radio luminosity function]{yua16}. In this section, we consider the effect of spectral index distribution on estimating the disk--jet correlation. 

The unified radio object catalog \citep{kim08,kim14} combined five radio catalogs and the optical SDSS survey catalog. Thus, this combined catalog provides the flux information of NVSS sources at different frequencies. Since \autoref{eq:Pjet_wil99} is calibrated at 151~MHz, we need a lower frequency flux density information for NVSS sources. The unified catalog includes the catalog by the Westerbork Northern Sky Survey \citep[WENSS;][]{ren97} at 325~MHz. It covers the sky north of $\delta=29^\circ$ with a limiting flux of $\sim18$~mJy and a beam size of $54"\times54"\csc(\delta)$. The positional accuracy is $\lesssim1.5"$ for bright sources and $\lesssim5"$ for faint sources.

\begin{figure}[t]
 \begin{center}
  \includegraphics[bb=0 0 864 648,width=8.5cm]{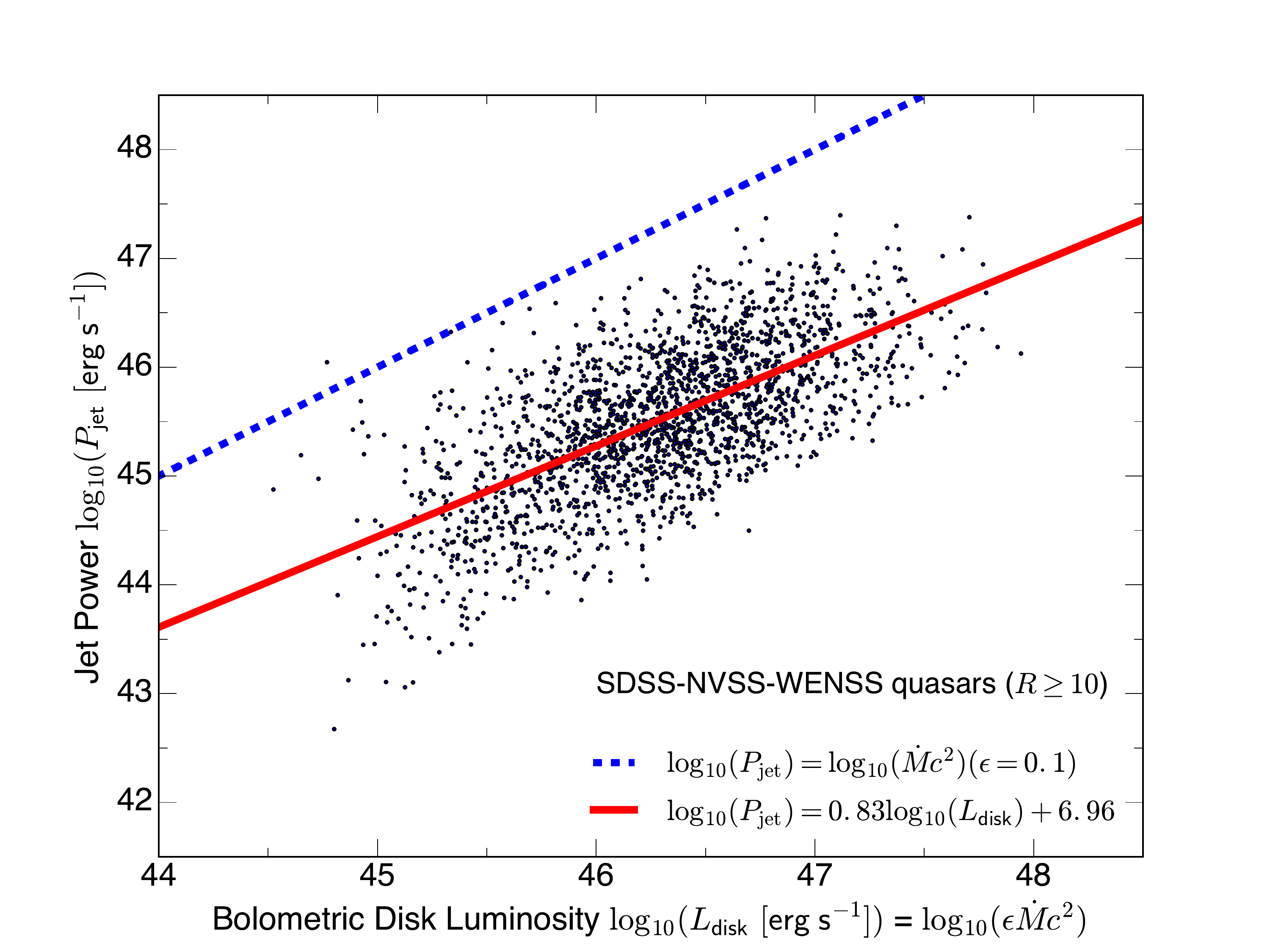} 
 \end{center}
\caption{Same as \autoref{fig:Ldisk_Pjet}, but for the SDSS--NVSS--WENSS detected quasar sub-sample.}\label{fig:Ldisk_Pjet_WENSS}
\end{figure}

The number of the SDSS-NVSS-WENSS detected radio-loud quasar sample is 2077 applying the same matching radius as our parent sample. We use the NVSS source position for the radio source position. Spectral index of individual object is estimated as
\begin{equation}
\alpha_{\rm NVSS-WENSS} = - \frac{\log(F_{\rm NVSS}/F_{\rm WENSS})}{\log{\rm (1.4~GHz/0.325~GHz)}},
\end{equation}
where we use the integrated flux density for radio fluxes. \autoref{fig:alpha_WENSS} shows the distribution of the spectral index. The distribution has $<\alpha_{\rm NVSS-WENSS}>=0.57\pm 0.50$, while we assume $\alpha=0.8$ for the main sample.

\autoref{fig:Ldisk_Pjet_WENSS} shows the correlation between $L_{\rm disk}$ and $P_{\rm jet}$ of our SDSS-NVSS-WENSS detected quasar sub-sample in the logarithmic space. The regression line, shown by the solid line, is given as
\begin{equation}
\log P_{\rm jet} = (0.83 \pm 0.020) \log L_{\rm disk} + (7.0 \pm 0.94),
\end{equation}
with a scatter of 0.51. The Spearman rank correlation gives $\rho_{LP,z}=0.31$ with the p-value of $<10^{-10}$. Therefore, there is still a positive correlation between disk luminosity and jet power in radio quasars. Based on the jet production efficiency arguments above, here we have $<\log \eta_{\rm jet}>=-1.86\pm0.53$ assuming $\epsilon=0.1$. Therefore, our results are not changed even if we take into account individual source spectral index variation.

\begin{figure}[t]
 \begin{center}
  \includegraphics[bb=0 0 864 648,width=8.5cm]{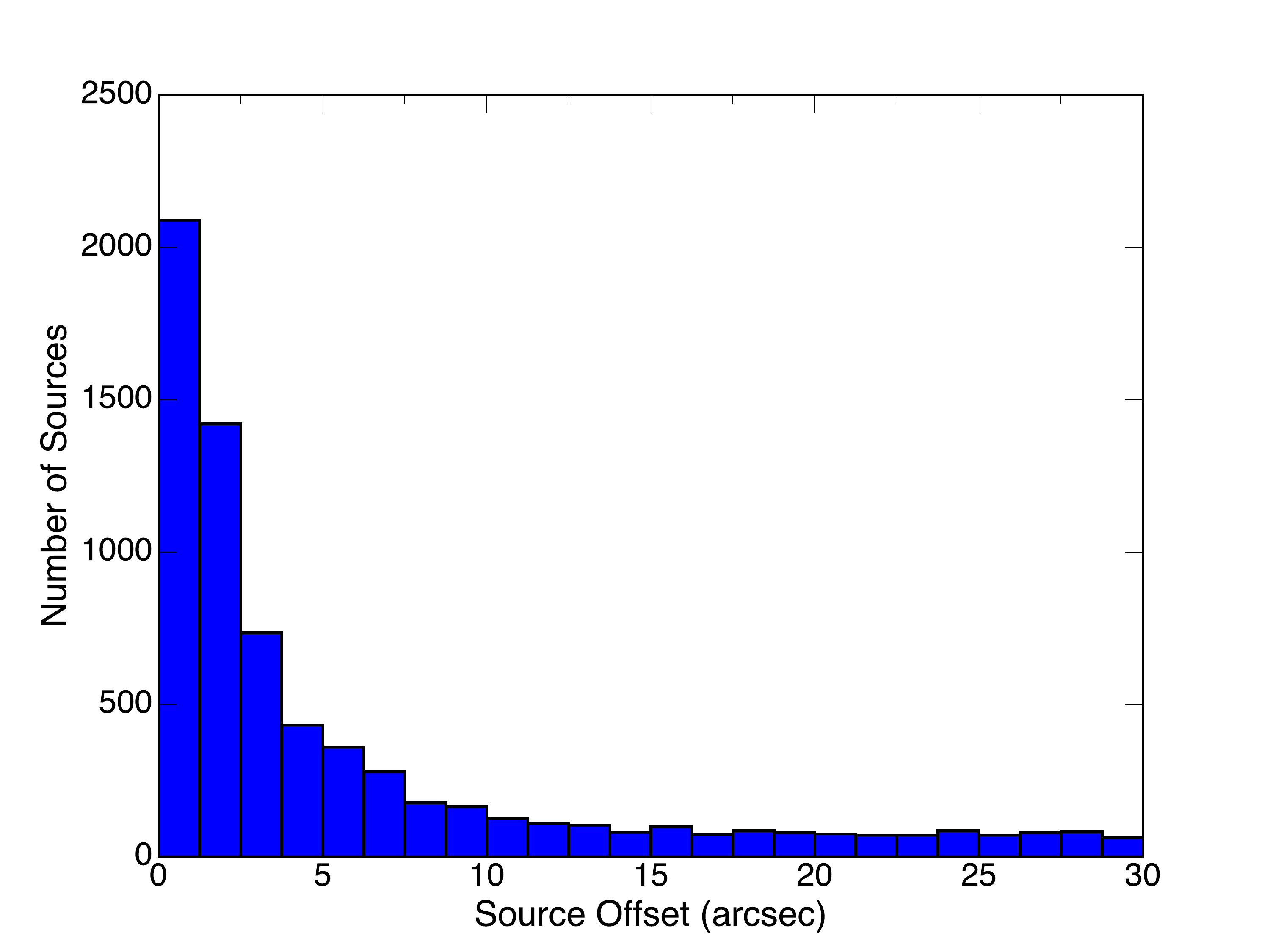} 
 \end{center}
\caption{Source offset distribution between the SDSS object and the NVSS object.}\label{fig:offset}
\end{figure}

\subsection{Different Sensitivity Limits}
The primary SDSS DR7 quasar catalog does not provide a uniform catalog due to a variety of criteria for target selection \citep{she11}.  Therefore, \citet{she11} provided a catalog flag representing the uniformity of the sample. ``Uniform flag = 1`` gives a uniformly selected quasar samples with a flux limit of $i=19.1$ at $z<2.9$ and $i=20.2$ at $z>2.9$ based on the target selection algorithm in \citet{ric02}. Faint source detection in the NVSS catalog may also suffers from such a completeness problem. Here, the 5-$\sigma$ detection sensitivity limit of NVSS is 2.5~mJy \citep[e.g.][]{con98,kim08}. 

To check such sample selection bias effects, we restrict our samples to the uniformly selected SDSS quasars with 5-$\sigma$ detection at the NVSS band. The resulting number of samples is 3545 radio-loud quasars. The relation between $P_{\rm jet}$ and $L_{\rm dis}$ is given as
\begin{equation}
\log P_{\rm jet} = (1.1\pm0.017)\log L_{\rm disk} + (-4.8 \pm 0.78),
\end{equation}
with a scatter of 0.53. The Spearman rank correlation gives $\rho_{LP,z}=0.37$ with the p-value of $<10^{-10}$. And, we have $<\log \eta_{\rm jet}>=-2.03\pm0.53$. Therefore, our results do not change even if we take into account the uniformity of the sample.

\subsection{Matching Radius}
\label{subsec:offset}
In this paper, we match the NVSS radio sources and the DR7 quasar catalog with a matching radius of 30~arcsec and select the closest sources between the catalogs. However, unrelated objects may be matched in our catalog. \autoref{fig:offset} shows the distribution of the offset between the NVSS and SDSS positions. 84\% of objects have the offsets smaller than 15~arcsec. The fraction becomes $\gtrsim$90\% for the offsets of $\le$20~arcsec. Therefore, the fraction of the mismatched objects would be minor in our parent catalog \citep{bes05}.

We restrict our samples to the SDSS--NVSS quasars with the matching radius of 15~arcsec to consider the selection effect by the matching radius. This sub-sample contains 6081 radio-loud objects. The relation between $P_{\rm jet}$ and $L_{\rm dis}$ with the matching radius of 15~arcsec is given as
\begin{equation}
\log P_{\rm jet} = (0.94\pm0.013) \log L_{\rm disk} + (1.8 \pm 0.60),
\end{equation}
with a scatter of 0.54. The Spearman rank correlation gives $\rho_{LP,z}=0.39$ with the p-value of $<10^{-10}$.  Therefore, the correlation between disk luminosity and jet power still statistically exists. The jet production efficiency is given as $<\log \eta_{\rm jet}>=-1.93 \pm0.54$.

\begin{figure}[t]
 \begin{center}
  \includegraphics[bb=0 0 864 648,width=8.5cm]{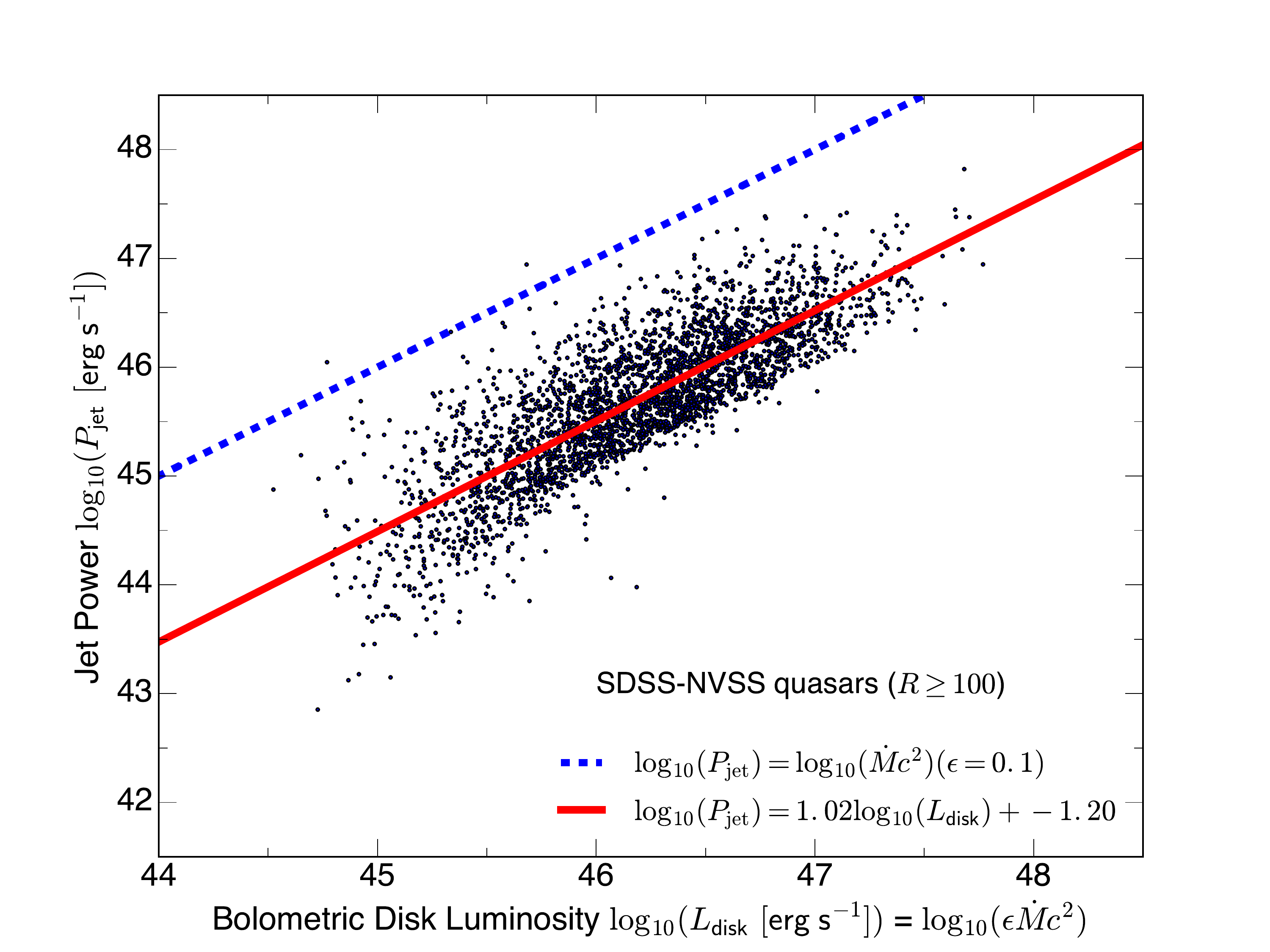} 
 \end{center}
\caption{Same as \autoref{fig:Ldisk_Pjet}, but $R\geq100$.}\label{fig:Ldisk_Pjet_logR2}
\end{figure}

\subsection{Effect of Radio Loudness Cut}
At $10\leq R\lesssim100$, radio-loud quasars are also called as radio-intermediate quasars \citep{fal96}. Most of them are likely to be FR-I type radio quasars. \autoref{fig:Ldisk_Pjet_logR2} shows the correlation between $L_{\rm disk}$ and $P_{\rm jet}$ of the SDSS-NVSS detected quasars but  having $R\geq100$. The dashed line represent $P_{\rm jet} =\dot{M}_{\rm in} c^2$. 3083 quasars are included. The regression line, shown by the solid line, is given as
\begin{equation}
\log P_{\rm jet} = (1.0 \pm 0.013) \log L_{\rm disk} + (-1.2 \pm 0.59),
\end{equation}
with a scatter of 0.38. The Spearman rank correlation gives $\rho_{LP,z}=0.63$ with the p-value of $<10^{-10}$. There is still a positive correlation between disk luminosity and jet power in radio quasars. Based on the jet production efficiency arguments above, here we have $<\log \eta_{\rm jet}>=-1.49\pm0.38$. By selecting sources at $R\geq100$, a slightly higher jet production efficiency is expected but still be consistent with that at $R\geq10$ at the 1-$\sigma$ level.

\begin{figure}[t]
 \begin{center}
  \includegraphics[bb=0 0 360 252,width=8.5cm]{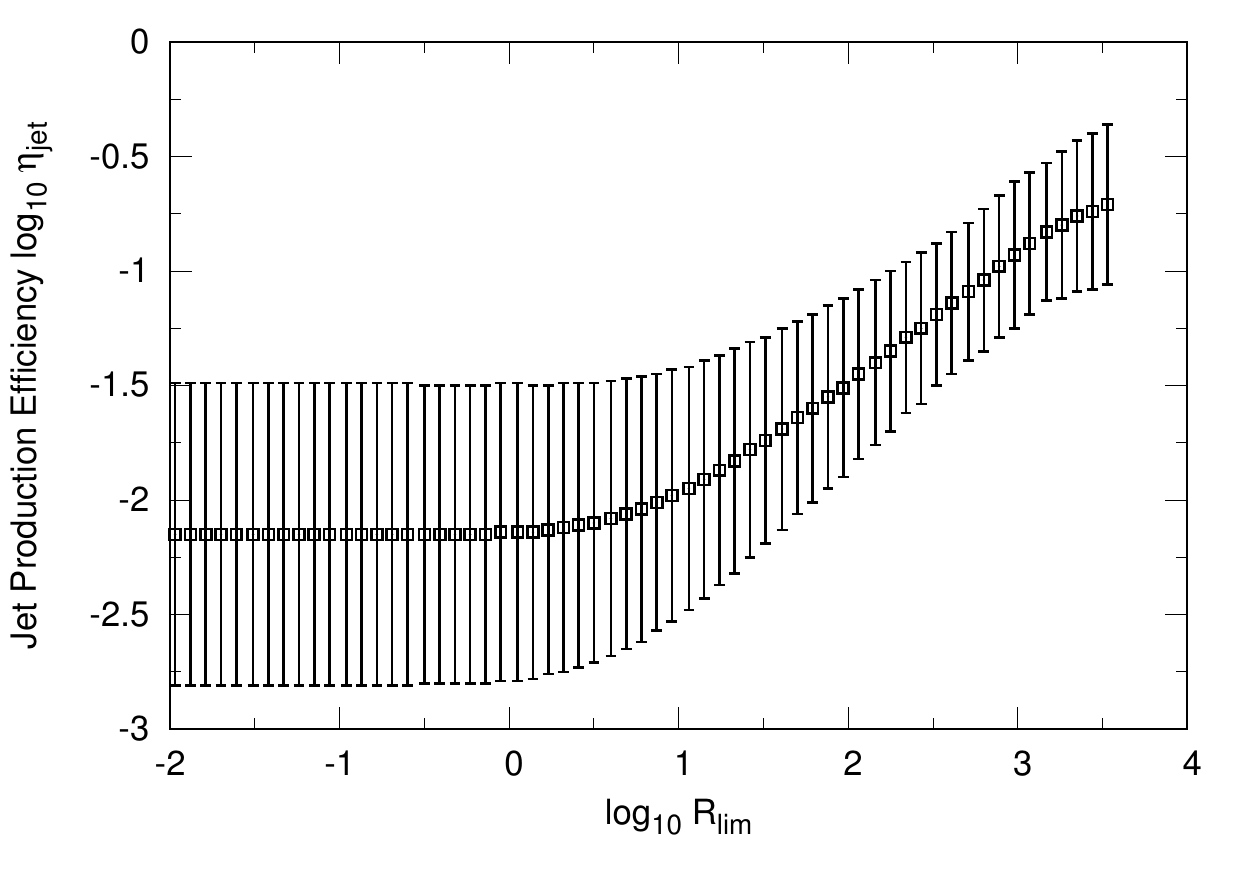} 
 \end{center}
\caption{Jet production efficiency $\eta_{\rm jet}$ as a function of the limiting radio loudness $R_{\rm lim}$. 1-$\sigma$ statistical uncertainty of the data is also shown.}\label{fig:eta_logR}
\end{figure}

\autoref{fig:eta_logR} shows the estimated jet production efficiency as a function of the limiting radio loudness. As we select high jet power objects by setting high $R_{\rm lim}$, the jet production efficiency gradually increases with $R_{\rm lim}$. However, even at $R_{\rm lim}\gtrsim10^3$, $\eta_{\rm jet}$ is still at an order of 0.1.

\subsection{Comparison with Blazar Studies}
The relation between disk and jet has been also investigated using blazar spectral fitting studies \citep[e.g.][]{cel08,ghi14,ino16}. Different from our radio quasar studies, blazar studies enable us to compare accretion with jet outflow at the same epoch. Multi-wavelength observations from radio to gamma-ray allow us to study overall SED and physical parameters of jets via spectral fitting. A spectrum of a blazar is composed of two emission components:  the low-energy component is synchrotron radiation and the other one is inverse Compton component targeting synchrotron radiation \citep[e.g.][]{jon74,mar92} and/or external radiation field \citep[e.g.][]{der93,sik94}. 

Generally, blazar spectral fitting studies predict large jet production efficiencies $\eta_{\rm jet}\sim$0.1--10 \citep[e.g.][]{cel08,ghi14,ino16}, while our results show $\eta_{\rm jet}\simeq1.1_{-0.76}^{+2.6}\times10^{-2}$. Such a difference has also been recently reported by \citet{pja16} using published blazar samples. Here, the jet power estimates based on blazar spectral fittings strongly rely on the assumptions on the pair fraction and the electron minimum Lorentz factor $\gamma_{e, \rm min}$ \citep[see e.g.][]{ino16,pja16}. More pairs per one proton  and/or higher $\gamma_{e, \rm min}$ would reduce the jet power estimation based on blazar spectral fit. Although $\gamma_{\rm min}$ is known to be $\sim m_p/m_e$ for the terminal shocks of quasar jets \citep{sta07}, the minimum Lorentz factor of electrons in blazar jet is not well constrained. Recent {\it NuSTAR} observations revealed that electron spectrum extends down to electron energies of at least $\gamma_e\sim10^2-10^3$ for nearby BL Lacs \citep{kat16,mad16}. If $\gamma_{\rm min}$ is determined by the mass ratio of a proton and an electron (i.e. $\gamma_{\rm min}=m_p/m_e$), the jet power can be an order of one to two lower than the case in which $\gamma_{\rm min}$ is determined by the energy ratio between protons and electrons \citep{ino16}, which would lead a consistent result with our radio quasar studies.

\subsection{Implication on to the Jet Production}
\label{subsec:impjet}

Recent general relativistic magnetohydrodynamic (GRMHD) numerical studies showed that powerful relativistic jets are launched in the MAD which can confine magnetic flux on BHs by its ram pressure \citep{tch11,mck12}. Extracted jet power by the rotation of BHs threaded by magnetic fields, so-called the BZ power, is given by \citep{bla77,tan08,tch10,tch11}
\begin{eqnarray}
P_{\rm BZ} &=& 4.0\times10^{-3}\frac{1}{c}\Omega_{\rm H}^2 \Phi_{\rm BH}^2f(\Omega_{\rm H})\\
&\simeq& 10 \left(\frac{\phi_{\rm BH}}{50}\right)^2x_a^2f(x_a)\dot{M}_{\rm in}c^2,
\end{eqnarray}
where $\Omega_{\rm H}=ac/2r_{\rm H}$ is the angular frequency of the BH horizon, $\Phi_{\rm BH}$ is the net magnetic field flux accumulated in the central region, $x_a\equiv r_g\Omega_{\rm H}/c$, and $f(x_a)\approx1+1.38x_a^2-9.2x_a^4$. $a\equiv J_{\rm BH}/J_{\rm BH, max}=cJ_{\rm BH}/GM_{\rm BH}^2$ is the dimensionless BH spin parameter, $r_{\rm H}=r_g(1+\sqrt{1-a^2})$ is the horizon radius, $r_g = GM_{\rm BH}/c^2$ is the gravitational radius of the BH. $\phi_{\rm BH}=\Phi_{\rm BH}/\sqrt{\dot{M}_{\rm in}r_g^2c}$ is the dimensionless magnetic flux threading the BH and is typically on the order of 50 based on GRMHD simulations \citep{mck12}\footnote{In \citet{mck12}, the dimensionless magnetic flux is denoted by $\Upsilon_{\rm BH}\approx \Phi_{\rm BH}/5\sqrt{\dot{M}_{\rm in}r_g^2c}=\phi_{\rm BH}/5$ \citep{gam99,pen10}. Typically, $\Upsilon_{\rm BH}$ is the order of 10 \citep[see Table 9 in][]{mck12}.}. This gives $\eta_{\rm jet, BZ}\equiv P_{\rm BZ}/\dot{M}_{\rm in}c^2 \simeq10 (\phi_{\rm BH}/50)^2x_a^2f(x_a)$. 

\begin{figure}[t]
 \begin{center}
  \includegraphics[bb=0 0 864 648,width=8.5cm]{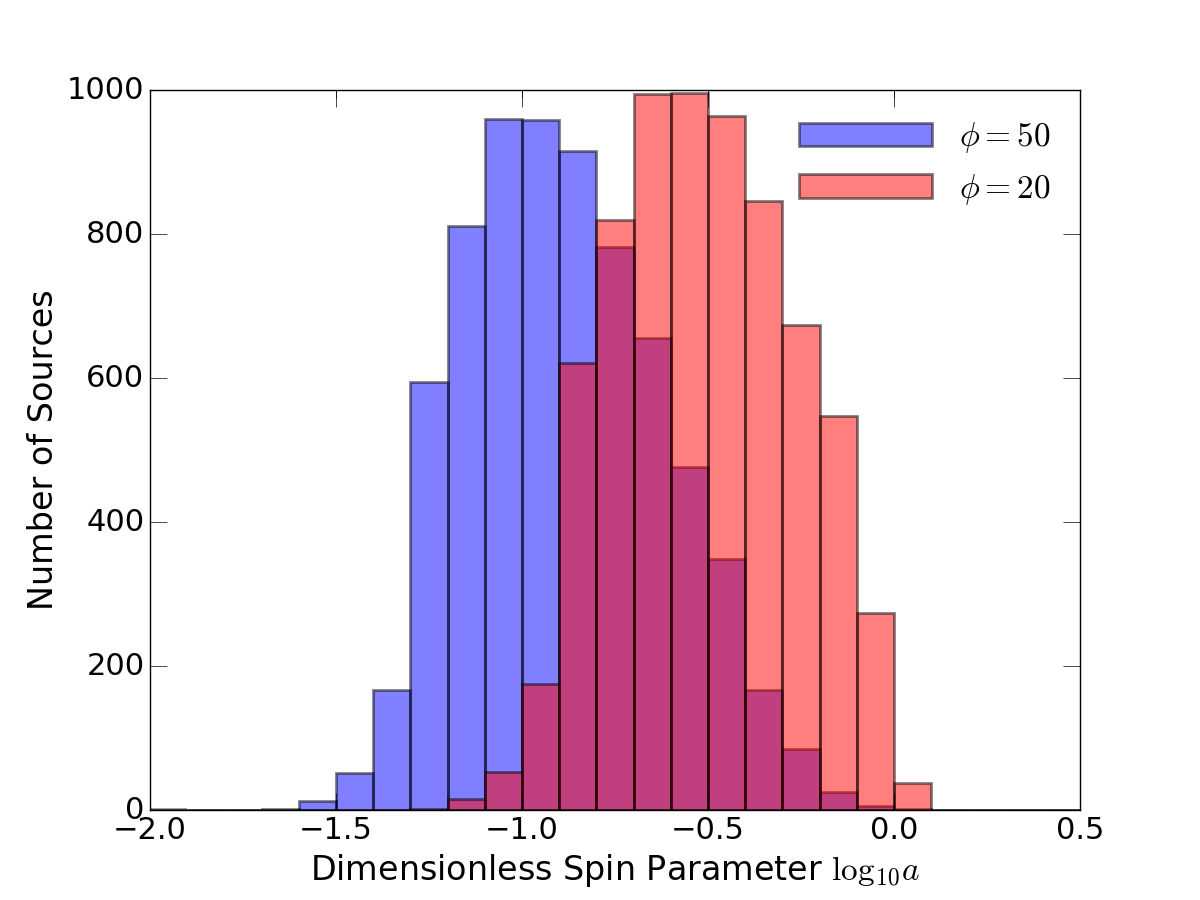} 
 \end{center}
\caption{Distribution of the dimensionless spin parameters of our radio quasar samples. The dimensionless magnetic flux $\phi_{\rm BH}$ is set to be 50 (blue) and 20 (red). We assume the radiative efficiency of the disk $\epsilon$ to be 0.1.}\label{fig:hist_loga}
\end{figure}

\autoref{fig:hist_loga} gives the distribution of the dimennsionless spin parameter $a$ assuming $\epsilon=0.1$ and $\phi_{\rm BH}=50$. For $\phi_{\rm BH}=50$, $<\log a>=-0.88\pm0.26$ corresponding to $a\simeq0.13_{-0.059}^{+0.11}$. However, this spin parameter value is smaller than what is expected from the cosmological merger SMBH evolution models  \citep{vol05,vol07,vol13}. \autoref{fig:hist_loga} also shows the spin parameter distribution but assuming $\phi_{\rm BH}=20$. In this case, we have $<\log a> = -0.50 \pm 0.24$. Therefore, the results on the spin parameter strongly depend on the assumption on the magnetic flux threading the BH which is expected to be the dominant factor for the generation of jets considering the radio loundness distribution \citep{sik13_rad}. 

Observationally, various attempts have been recently considered to measure the magnetic field near active SMBHs. VLBI observations toward M~87, whose accretion rate is significantly sub-Eddingtion  \citep{dim03}, revealed $B\sim10$~G at the jet base $\sim10R_s$ from the central BH using the synchrotron self-absorption (SSA) frequency \citep{kin15,had16}. Recent linear polarimetric adaptive optics observation of an AGN torus yields a dusty torus magnetic filed strength in the range of 4--82~mG assuming a clumpy torus model \citep{lop15}. Maser observations also constrains the magnetic fields at $\sim0.2$~pc from the central BH to be $\lesssim1$~G \citep{gne14}. Although the B-field in the inner region (near the corona scale) has not been well investigated yet, future detection of coronal synchrotron emission would provide an information on magnetic field \citep{ino14,rag16}. Possible millimeter excess has been already reported from a nearby radio-quiet AGN NGC~985 \citep{doi16}. If this excess is coming from the corona, the magnetic field strength at 100 times the Schwarzschild radius is expected to be $\sim150$~G \citep{doi16} near the equipartition value with the gas energy density. Assuming $\epsilon=0.1$, this yields $\phi_{\rm corona}\sim2$. If $\phi_{\rm BH}$ is such a low value even for radio quasars, faster spin parameters are required. 

\citet{ava16} have recently found that the jet production efficiency in the MAD scenario depend not only on the spin and magnetic flux but also on the geometrical thickness of the accretion flow. Geometrical thickness of the accretion from of our quasar samples is expected to be small since they are at the range of $0.01 \lesssim \lambda_{\rm disk} \lesssim 0.3$ (See \autoref{fig:lambda-R}). However, the disk structure itself is not well understood. In the MAD scenario, the disk is expected to be truncated at the outer radius of a magnetospheric radius \citep{nar03}. Although we would be able to locate the inner radius based on Iron-K line analysis \citep{ger91}, it is still not clear whether there is any systematic trend of disk truncation in radio-loud AGNs. Some radio-loud AGNs show the inner radius is $>10 R_s$ \citep[e.g.,][]{lar08,sam09,taz10,tom11,taz13}, while the disks of others extend down to $\sim4-6R_s$ \citep{kat07,sam11}.

\subsection{Radio Loudness and Disk Eddington Ratio}
\label{sec:R-lambda}

The correlation between $\lambda$ and $R$ (\autoref{fig:lambda-R}) is also another important indicator for the disk and jet connection. 
\citet{sik07} found a negative correlation between the radio loudness $R$ and the Eddington accretion ratio by covering nearly 7 orders of magnitude in disk Eddington ratios. And, a clear bimodality is seen in the correlation \citep{sik07}, although \citet{bro11} pointed out that the bimodality becomes less apparent when $R$ is determined from the core luminosity only and a mass correction is applied \citep[see also][]{gar14}. The physical origin of the correlation between $\lambda$ and $R$ is still under debate. 

The correlation can be tested with our full quasar sample (\autoref{fig:lambda-R}), although our sample covers only about 3 orders of magnitude in disk Eddington ratios ($0.01\lesssim\lambda\lesssim1$). The Spearman rank correlation gives $\rho_{\lambda R}=-0.11$ with the p-value of $<10^{-10}$. Therefore, a negative correlation between $\lambda$ and $R$ exists at $0.01\lesssim\lambda\lesssim1$. Here, both $R$ and $\lambda$ depend on the disk luminosity. If $L_{\rm disk}$ is set as the control variable, $\rho_{\lambda R, L}$ becomes -0.022. 

This weak negative correlation between $R$ and $\lambda$ can be fitted by the function of
\begin{equation}
\log R = (-0.28\pm0.015) \log \lambda + (1.5 \pm 0.015),
\end{equation}
with a scatter of 0.74. Following the MAD scenario \citep{nar03}, it is expected to be $R\propto \eta_{\rm jet}/\epsilon\propto\lambda^{-0.4}$ \citep{sik13}. Although we find $R\propto\lambda^{-0.28}$, distributions of $M_{\rm BH}$ and magnetic fluxes would cause dispersion on this relation. \citet{sik13} argued that $R$ can decrease with increasing $\lambda$ if all cold accretion episodes start after the magnetic flux accumulation by a hot (geometrically thick) accretion phase. Indications of such a process during the Bondi accretion phase have been recently provided by studies of $P_{\rm jet}/\dot{M}_{\rm Bondi}$ in nearby radio galaxies \citep{nem15}. \citet{rus16} proposed that time variation would cause this trend. The jet power estimated in this paper is time-averaged values, while the disk luminosity is the instantaneous value as discussed above. Since our sample covers only $0.01\lesssim\lambda\lesssim1$, additional samples covering lower $\lambda$ are necessary for more detailed analysis on this relation.

\section{Conclusions}
\label{sec:con}
In this paper, we studied the disk-jet connection in active SMBHs utilizing 7017 SDSS-NVSS detected radio-loud quasars up to $z=4.9$. We converted the 1.4~GHz radio luminosity to the jet power using an empirical relation \citep{wil99} which is calibrated by the X-ray cavity measurements \citep{wil99,god13}. Bolometric accretion disk luminosity is estimated by the SED templates in \citet{ric06}. Central BH mass is also provided by \citet{she11} using a single-epoch spectrum. 

We found that the quasar jet powers correlate with the bolometric disk luminosities. We have $\log P_{\rm jet} = (0.96\pm0.012)\log L_{\rm disk} + (0.79 \pm 0.55)$ with a scatter of 0.54. The jet power rarely exceeds the accretion luminosity. By assuming the accretion disk efficiency of $\epsilon=0.1$, we further found that the jet production efficiency is $\eta_{\rm jet}\simeq1.1_{-0.76}^{+2.6}\times10^{-2}$. These results do not significantly change even if we adopt various different selection criteria or  the higher disk efficiency of $\epsilon=0.3$. $\eta_{\rm jet}$ gradually increases with the limiting radio loudness, although it will be still at an order of 0.1 even for $R_{\rm lim}\gtrsim10^3$. 

We further tested the existence of the fundamental plane among $M_{\rm BH}$, $L_{\rm disk}$, and $P_{\rm jet}$ for quasars at $0.01\lesssim\lambda$. We could not find a statistically significant correlation between $M_{\rm BH}$ and $P_{\rm jet}$ excluding the dependence on $L_{\rm disk}$. This implies that the plane would not exist for radio-loud quasars in the standard accretion regime. This is consistent with recent studies which revealed that the plane exists only for low accretion AGNs \citep[e.g.][]{mer08,plo12}.

The relation between radio loudness, $R$, and the disk Eddington ratio $\lambda$ is also investigated. Our samples cover about 3 orders of magnitude in $\lambda$. With our large sample, we confirmed that there is a weak negative correlation between $R$ and $\lambda$. This is consistent with the report in \citet{sik07}. However, our samples covered only at $10^{-2}\lesssim\lambda\lesssim10$. Future deeper optical spectroscopic surveys such as the Subaru Prime Focus Spectrograph survey \citep{tak14,sug15} would help us to extend our samples to optically faint AGNs, i.e. high radio-loudness objects.

Following the BZ scenario \citep{bla77}, the jet power is related to the BH spin and the magnetic flux threading the BH. Taking the value of the dimensionless magnetic flux value from numerical simulations by \citet{mck12} $\phi_{\rm BH}=50$, we estimated the distribution of the dimensionless spin parameters $a$. However, the resulting $a$ is $\simeq0.13_{-0.059}^{+0.11}$ which is much smaller than what is expected from the cosmological merger SMBH evolution \citep[e.g.,][]{vol13}. Therefore, the magnetic flux strength threading the SMBHs might be weaker than it is expected from numerical simulations following the spin evolutionary models. Magnetic field measurements in the vicinity of the central BHs would help us to understand the detailed BZ processes. The resulting low jet efficiency can be also due to dependence of the efficiency on geometrical thickness of the accretion flow \citep{ava16}, since our quasar samples accreting at $0.01 \lesssim \lambda_{\rm disk} \lesssim 0.3$ is expected to be have small geometrical thickness.

\bigskip
The authors thank Chris Done, Hirokazu Odaka, and Andreas Schulze for useful comments and discussions. The authors also thank the anonymus referee for his/her useful and constructive comments which improved the paper a lot. YI is supported by the JAXA international top young fellowship and JSPS KAKENHI Grant Number 420 JP16K13813. STSDAS is a product of the Space Telescope Science Institute, which is operated by AURA for NASA. 

Funding for the SDSS and SDSS-II has been provided by the Alfred P. Sloan Foundation, the Participating Institutions, the National Science Foundation, the U.S. Department of Energy, the National Aeronautics and Space Administration, the Japanese Monbukagakusho, the Max Planck Society, and the Higher Education Funding Council for England. The SDSS Web Site is http://www.sdss.org/. The SDSS is managed by the Astrophysical Research Consortium for the Participating Institutions. The Participating Institutions are the American Museum of Natural History, Astrophysical Institute Potsdam, University of Basel, University of Cambridge, Case Western Reserve University, University of Chicago, Drexel University, Fermilab, the Institute for Advanced Study, the Japan Participation Group, Johns Hopkins University, the Joint Institute for Nuclear Astrophysics, the Kavli Institute for Particle Astrophysics and Cosmology, the Korean Scientist Group, the Chinese Academy of Sciences (LAMOST), Los Alamos National Laboratory, the Max-Planck-Institute for Astronomy (MPIA), the Max-Planck-Institute for Astrophysics (MPA), New Mexico State University, Ohio State University, University of Pittsburgh, University of Portsmouth, Princeton University, the United States Naval Observatory, and the University of Washington.
\vspace{5mm}
\facilities{NVSS, SDSS, WENSS}

\appendix
\section{Correlation Analysis with Censored Data}
\label{app:corr}
In this paper, we use only detected data. However, our samples are flux-limited samples. Given the SDSS DR7 quasar catalog, a portion of those quasars are not detected in NVSS. Such a censored data set would affect the results and implication to the data \citep[e.g.][]{iso86,bon13}. In this section, we argue how the results will be affected by such a censored data.

\begin{figure}[t]
 \begin{center}
  \includegraphics[bb=0 0 360 252,width=8.5cm]{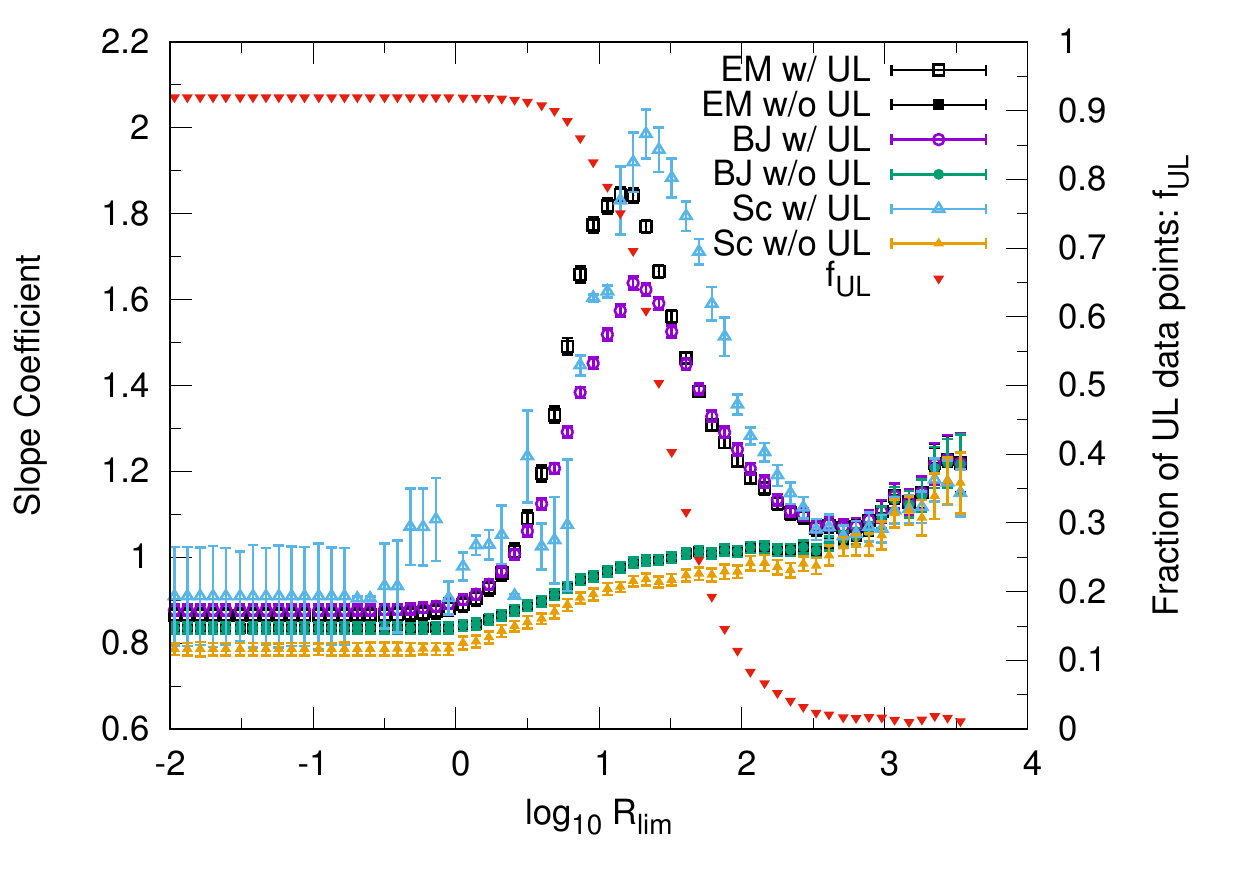} 
 \end{center}
\caption{Slope coefficient of the regression line between $L_{\rm disk}$ and $P_{\rm jet}$ as a function of limiting radio loudness $R_{\rm lim}$. Open data treat only detected samples, while filled data include censored data. Square, circle, and upper-triangle point corresponds to the results from the expectation-maximization (EM) algorithm, the Buckley-James (BJ) algorithm, and the Schmitt's (Sc) algorithm, respectively. The down-triangle points show the number fraction of non-detected samples.}\label{figa:slope}
\end{figure}

\begin{figure}[t]
 \begin{center}
  \includegraphics[bb=0 0 360 252,width=8.5cm]{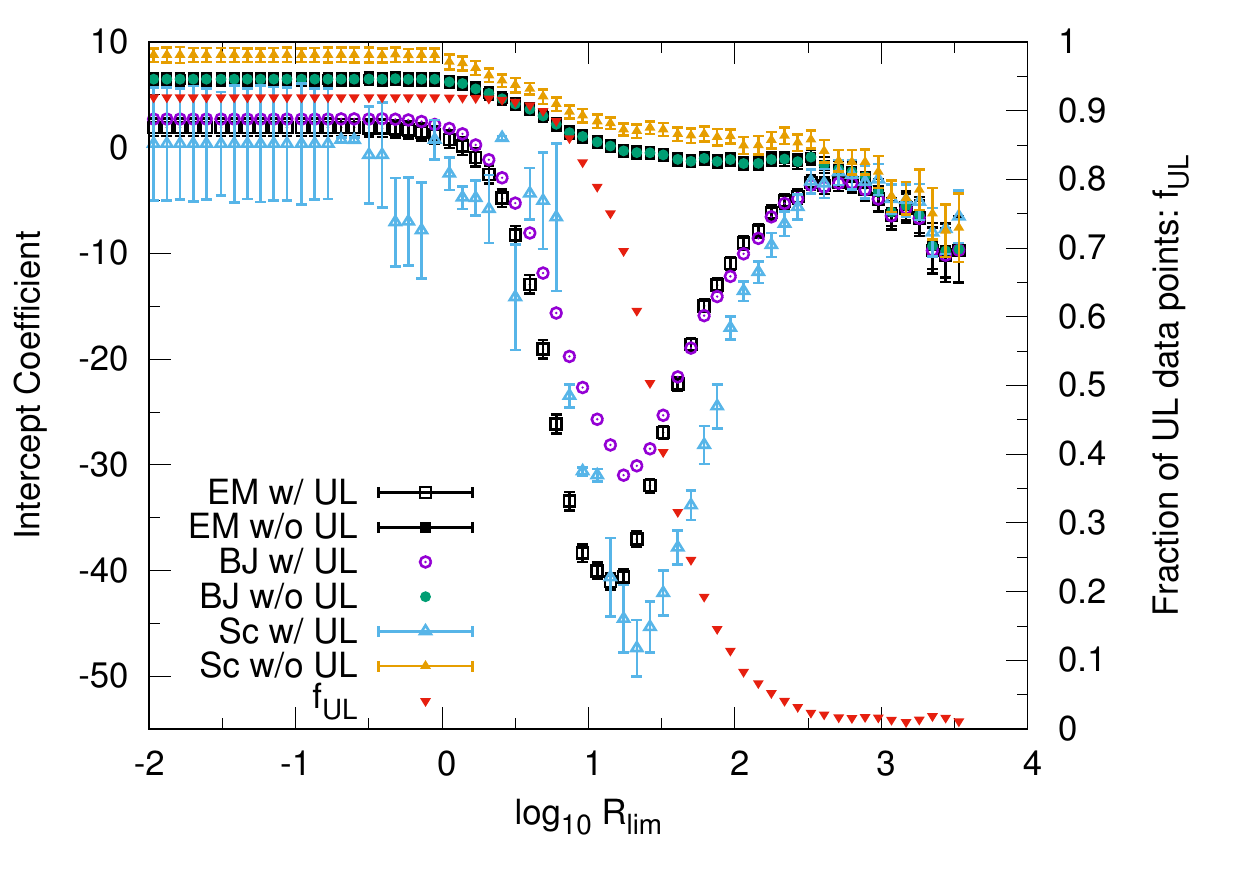} 
 \end{center}
\caption{Same as \autoref{figa:slope}, but showing intercept coefficients.}\label{figa:intercept}
\end{figure}

The number of the SDSS DR7 quasars is 105783 \citep{she11}, while our parent sample contains 8436 SDSS-DR7 quasars of which 7017 are radio-loud. Since it is known that only about 10\% of quasars are radio-loud \citep{bal12}, this large difference is naturally expected. Therefore, if we include all the SDSS DR7 quasar, radio data are dominated by the upper limits of unrelated radio-quiet objects whose radio emission is generated by quasar driven disk outflows rather than relativistic jets \citep[e.g.][]{zak14}. In order to investigate radio-loud objects, we restrict samples above a certain radio-loudness $R_{\rm lim}$. For non-detected objects, we can select sources by setting the upper limits on the radio loudness $R_{\rm UL}$ above $R_{\rm lim}$, where radio flux upper limits for non-detected objects are set as the detection limit of the NVSS, $2.5~{\rm mJy}$ \citep{con98}.

\begin{figure}[t]
 \begin{center}
  \includegraphics[bb=0 0 864 648,width=8.5cm]{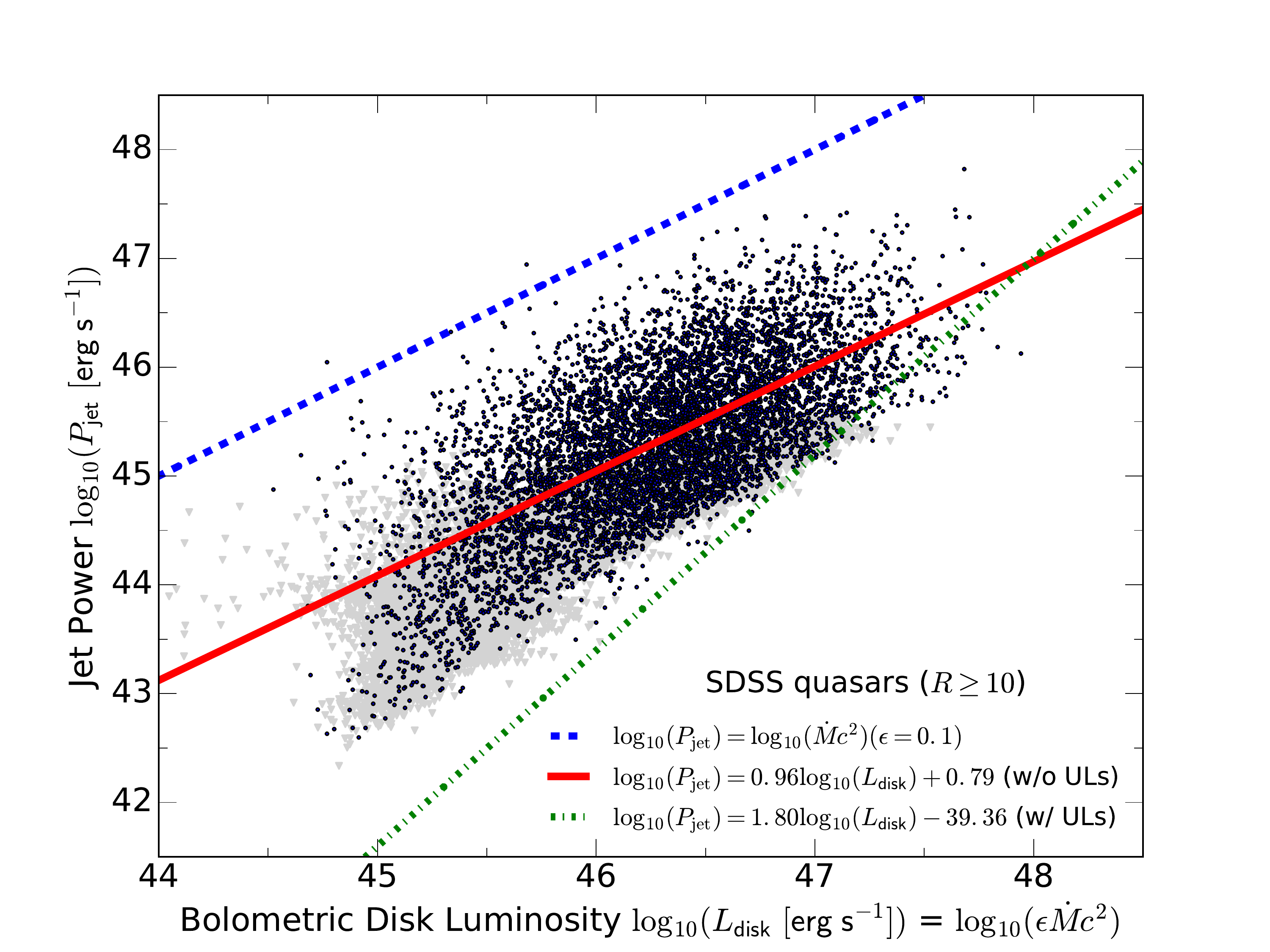} 
 \end{center}
\caption{Same as \autoref{fig:Ldisk_Pjet}, but including censored data (gray triangles). Regression line including the censored data is shown in dot-dashed. The EM algorithm is adopted.}\label{figa:Ldisk_Pjet}
\end{figure}

\autoref{figa:slope} and \autoref{figa:intercept} show the slope coefficient and the intercept coefficient, respectively, in the regression analysis of the $(L_{\rm disk}, P_{\rm jet})$ plane as a function of $R_{\rm lim}$ for the data sets with and without non-detected objects. We adopt the three regression analysis methods; expectation-maximization (EM) algorithm, the Buckley-James (BJ) algorithm, and the Schmitt's (Sc) algorithm. The EM algorithm is similar to the least-squares fitting method, and assumes that the intrinsic residuals of $P_{\rm jet}$ are normally distributed in log space about the regression line for fixed values of $L_{\rm disk}$. Censored data are taken into account by determining the degree to which the upper limits are compatible with the assumed dispersion about the regression line. The BJ algorithm is similar in approach, but uses the Kaplan-Meier distribution for the residuals. The Sc method divides the $(x,y)$ plane into bins and uses the Kaplan-Meier distribution for the residuals. \citet{iso86} provide details about theses algorithms.

The three regression algorithms give similar results for the entire the $R_{\rm lim}$ range. However, it is clear that the results becomes significantly different by the treatment of non-detection data samples, specifically at $\log R_{\rm lim}\lesssim2.5$ where the fraction of non-detected data starts to increases. In order to see how the results are affected by censored data, \autoref{figa:Ldisk_Pjet} shows the relation between $L_{\rm disk}$ and $P_{\rm jet}$ setting $R\geq10$ including censored data. The regression line for the detected sources is given in \autoref{eq:jet_disk}, while the regression line incorporating the censored data is given in 
\begin{equation}
\log P_{\rm jet} = (1.80\pm0.017)\log L_{\rm disk} + (-39.4 \pm 0.81),
\end{equation}
with a scatter of 1.10. The values are from the EM method assuming a normal distribution for the residuals. It is clearly seen that the regression line for the censored data does not well reproduce the detected data.

In the fitting procedure in the methods above, regression lines for the censored data generally become lower than the original upper limit (UL) values. This is because censored data are incorporated determining the degree to which the upper limits are compatible with the assumed dispersion about the regression line. The expected true values of the censored data from the regression analysis could become lower than the UL values. Thus, the true value of $R$ of the UL data would become $<R_{\rm lim}$. However, we need to restrict samples having $R\geq R_{\rm lim}$ in order to select objects having relativistic jets. This inconsistency potentially leads spurious outcomes by taking into account the censored data. Thus, in order to avoid such results due to unrelated datasets, we do not include the censored data in our analysis.

\end{document}